\documentclass[prb,showpacs,twocolumn,floatfix]{revtex4-1}
\usepackage{graphicx} 
\usepackage{dcolumn}  
\usepackage{bm}       
\usepackage{amsmath}
\usepackage{amsfonts}
\usepackage{amssymb}
\usepackage{times}

\begin{document}

\title{
The role of potassium orbitals in the metallic behavior of K$_3$picene
} 

\author{G. Chiappe}
\author{E. Louis}
\affiliation{Departamento de F\'{\i}sica Aplicada, Unidad Asociada del
CSIC and
Instituto Universitario de Materiales, Universidad
de Alicante, San Vicente del Raspeig, 03690 Alicante, Spain.}

\author{A. Guijarro}
\affiliation{Departamento de Qu\'{\i}mica Org\'anica and
Instituto Universitario de S\'{\i}ntesis Org\'anica, Universidad
de Alicante, San Vicente del Raspeig, 03690 Alicante, Spain.}

\author{E. San-Fabi\'an}
\affiliation{Departamento de Qu\'{\i}mica F\'{\i}sica, Unidad Asociada del
CSIC and
Instituto Universitario de Materiales, Universidad
de Alicante, San Vicente del Raspeig, 03690 Alicante, Spain.}

\author{J.A. Verg\'es}
\affiliation{Departamento de Teor\'{\i}a y Simulaci\'on de Materiales,
Instituto de Ciencia de Materiales de Madrid (CSIC), Cantoblanco, 28049 Madrid,
Spain.}
\email{jav@icmm.csic.es}

\date{\today}

\begin{abstract}
Detailed electronic structure calculations of picene clusters doped by potassium
modeling the crystalline K$_3$picene structure show that while two
electrons are completely transferred from potassium atoms to the LUMO of
pristine picene, the third one remains closely attached to both material
components. Multiconfigurational analysis is necessary to show that many
structures of almost degenerate total energies compete to define the cluster
ground state. Our results prove that the $4s$ orbital of potassium should be included
in any interaction model describing the material.
We propose a quarter-filled two-orbital model as the most simple model capable of
describing the electronic structure of K-intercalated picene.
Precise solutions obtained by a development of Lanczos method
show low energy electronic excitations involving orbitals located at different
positions. Consequently, metallic transport is possible in spite of the
clear dominance of interaction over hopping.
\end{abstract}



\maketitle

\section{Introduction}

Doping of crystals formed by large polycyclic aromatic hydrocarbon (PAH)
molecules like
picene has originated a new class of superconducting materials of very
promising characteristics. The pioneering work of Mitsuhashi
{\it et al.}\cite{ mitsu2010} showed superconductivity of potassium-intercalated
picene at 7K or 18K depending on sample processing.
Later, other PAH crystals formed by coronene\cite{coroneno2011},
phenanthrene\cite{fenantreno2011},
and dibenzopentacene\cite{dibenzo2012} have also shown superconducting
properties according to the temperature behavior of magnetic susceptibility.
Doping with group 2 elements is also possible\cite{Srfenantreno2011}.
Very recently, zero resistivity in K-doped picene has been
observed\cite{ZeroResistivity}.

\begin{figure}
$\begin{array}{cc}
\includegraphics[width=0.5\columnwidth]{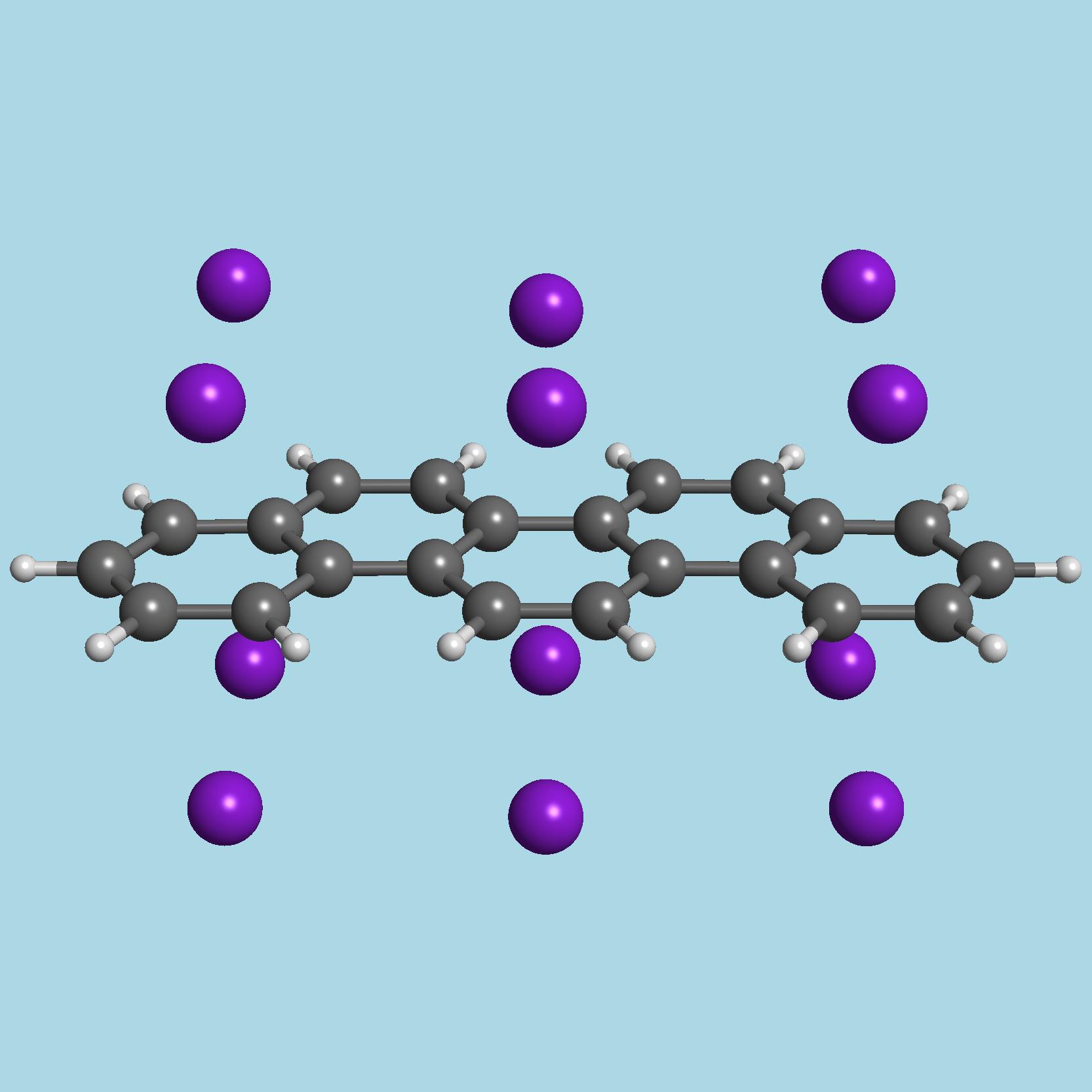} &
\includegraphics[width=0.5\columnwidth]{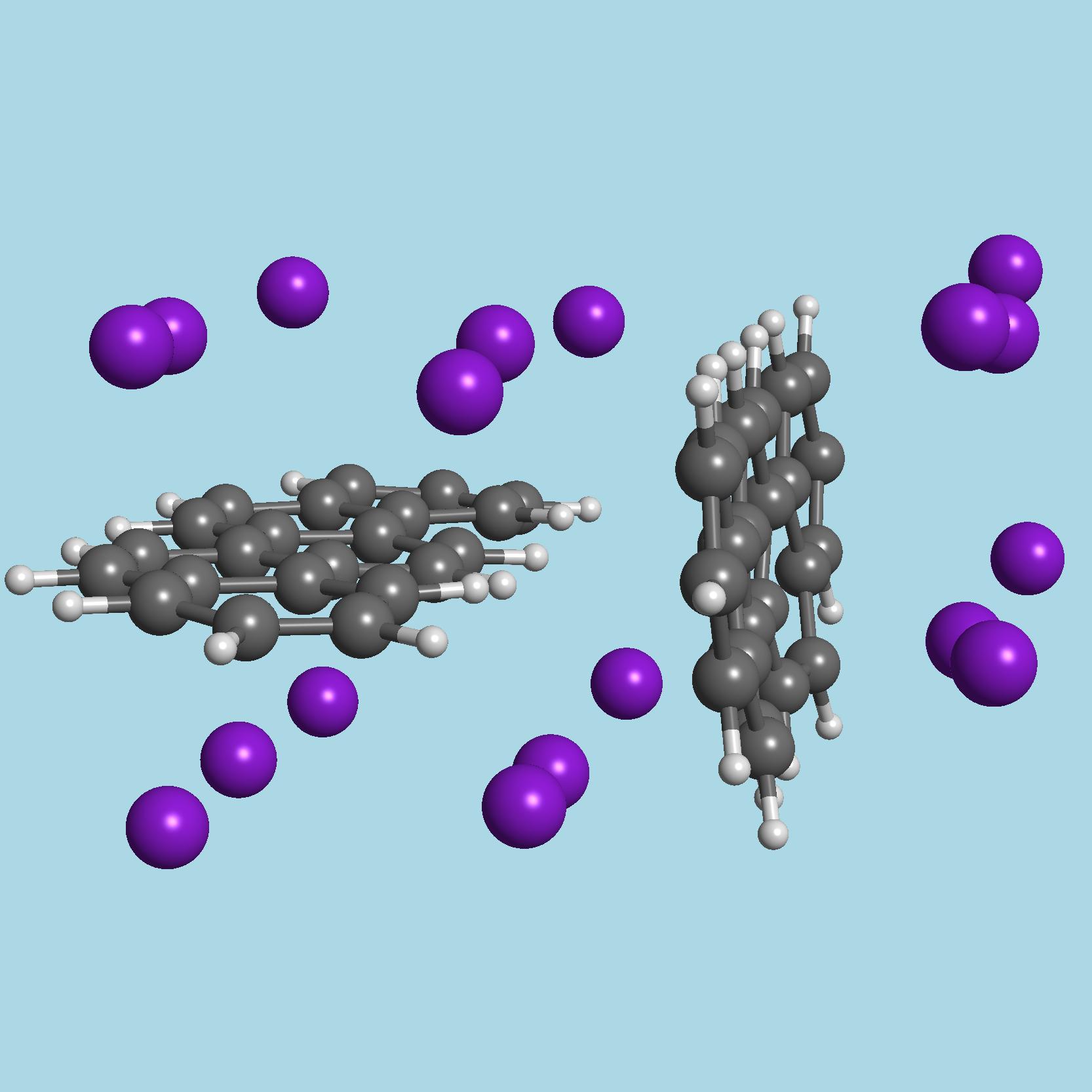} \\
\includegraphics[width=0.5\columnwidth]{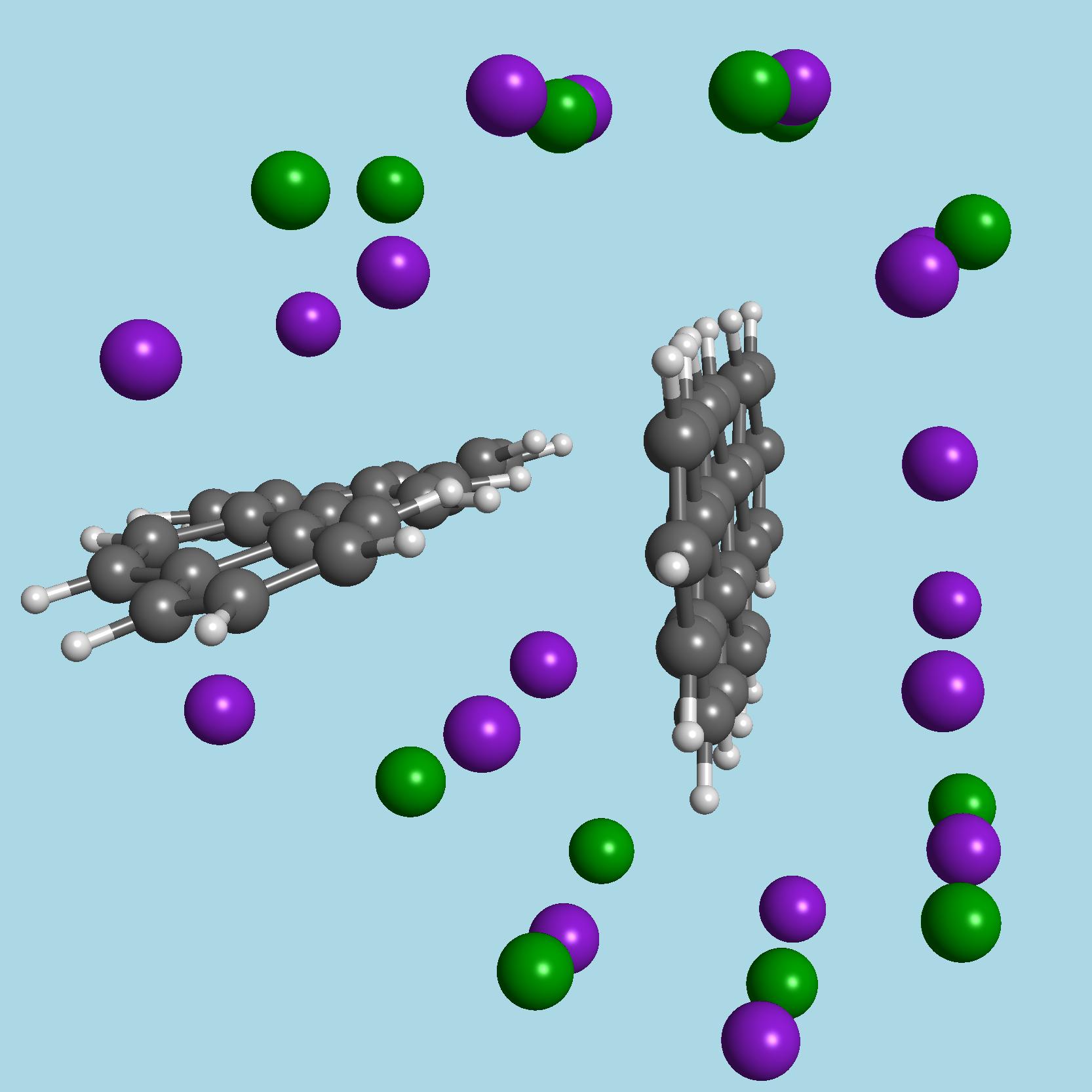} &
\includegraphics[width=0.5\columnwidth]{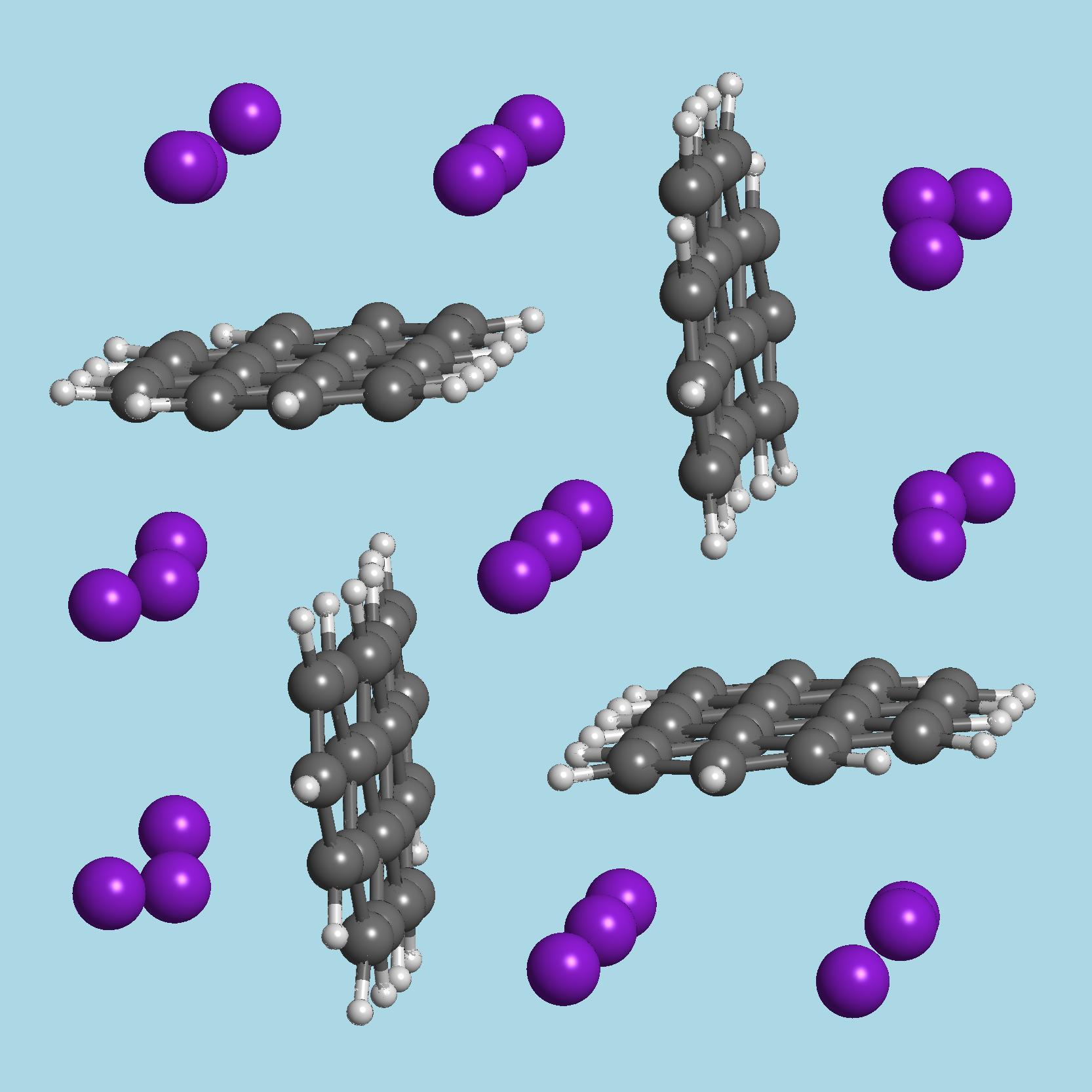} \\
\end{array}$
\caption{(Color online) Potassium-picene clusters used in the present
work. They are based in the probable crystalline structure of
K$_3$picene and have been fully optimized at the DFT level using
the B3LYP density functional. Numbering the figure panels from
left to right and from top to bottom,
the third one includes chlorine atoms to adjust
charge transference from potassium atoms to picene molecules.
}
\label{fig:clusters}
\end{figure}

Parallel to the experimental work there have been several theoretical
calculations of the electronic structure of potassium-intercalated picene\cite{
kos2009, giovannetti2011, pedro2011, kim2011, kosugi2011, kambe2012,
ruff2013, zhong2013}. According
to X-ray diffraction experiments, potassium enters into the intralayer
herringbone arrangement of picene molecules. The exact position of K atoms
in not known but several Density Functional Theory (DFT)
computations agree locating K atoms in
rows of three atoms that are approximately equidistant
to four picene molecules (see the central K group in the bottom right
panel of Fig. \ref{fig:clusters}).
In this way, the original structure of pristine
crystalline picene is minimally distorted. The relevant electronic bands are
above the original semiconducting gap of undoped picene. They are
modified by the presence of potassium and occupied by three electrons
coming from doping K atoms. It is usually assumed that the band formed
by the Lowest-energy Unoccupied Molecular Orbital (LUMO) of picene is completely
occupied whereas the LUMO+1 band remains half-occupied. This image
supports a conducting system that becomes superconducting when the
temperature decreases sufficiently. On the other hand, some papers have
pointed to a magnetic character of the last bands that invalidates the
naive picture of independent electrons\cite{giovannetti2011,kim2011,zhong2013,verges2012}.

In this paper, we want to emphasize the role of the $4s$ K orbital in
the {\em occupied} part of electronic band structure. While our
calculations show that two K electrons go to the picene LUMO, the third
K electron feels almost equal potentials at $4s$ atomic K orbital and picene
LUMO+1 and, consequently, remains in a kind of resonating situation\cite{IUPAC}.
We suggest that this capability of the last doping electron for
visiting both the PAH molecule and the doping site should be the
ultimate reason for the metallic behavior of the compound. 

The rest of the paper is organized as follows. Section II is devoted to
give some details of the methods and procedures used in this work.
The selection and geometry optimization of clusters is presented and
standard quantum chemistry methods valid in multi-configurational cases
are briefly described.
Section III gives our main quantum chemistry computational
results together with some discussion.
Section IV presents and discusses a minimal interaction model
for the most relevant electrons of doped picene.
It allows recovering the main results obtained for small
clusters. Besides, solutions for larger systems give
further support to our explanation of metallic charge transport.
The work ends with a few final concluding remarks (Section V).

\section{Computational procedures}

Potassium-doped picene clusters have been constructed based on the probable
crystalline structure of K$_3$picene\cite{xxx}.
The edge-to-face arrangement of picene molecules is similar to the one found in
pristine picene while the positions of K atoms are not
precisely known although all studies agree in having intralayer
potassium that is more or less regularly distributed between aromatic
molecules. In our study, we choose the most regular possible environments
for picene molecules always surrounding them by twelve K atoms
(see Fig. \ref{fig:clusters}). The price we
pay for this choice is a system that is too rich in potassium according to
nominal stoichiometry. Nevertheless, we have found a procedure that works
fine and correctly simulates the electronic structure of K$_3$picene.
Firstly, the neutral cluster is fully optimized at a Density Functional
Theory (DFT) level. Secondly, the number of electrons in the cluster is
reduced to make charge distribution as close as possible to the one
corresponding to the bulk material. In this way,
three electrons are allowed to flow to picene molecules while the rest
of K valence electrons are assumed to be transferred to more distant
molecules (that is, picene molecules that are not included in the cluster).
Actually, charge transference from $4s$ K orbital to LUMO and LUMO+1 or
LUMO+2 of picene is a bit
more subtle since the third electron is not completely transferred to picene
molecules. That means that K keeps part of its valence electron. After some
trial and error we have found that about one half of an electron remains
in the groups formed by three almost aligned potassium
atoms sitting between four picene molecules.
Consequently, picene molecules gain only two and a half electrons
and the total number of electrons of a cluster should be carefully modulated
to be consistent with this situation.
Prior to present detailed results, we would like to motivate the reader showing
precisely the charge distribution on K orbitals predicted by our approach
not only for the relatively small clusters that can be studied with high
numerical precision but also for larger clusters that have been studied
at a lower demanding computational cost
(see Fig. \ref{fig:SUPER} at the end of the Quantum Chemistry Section (III)). 

\begin{figure}
$\begin{array}{cc}
\includegraphics[width=0.49\columnwidth]{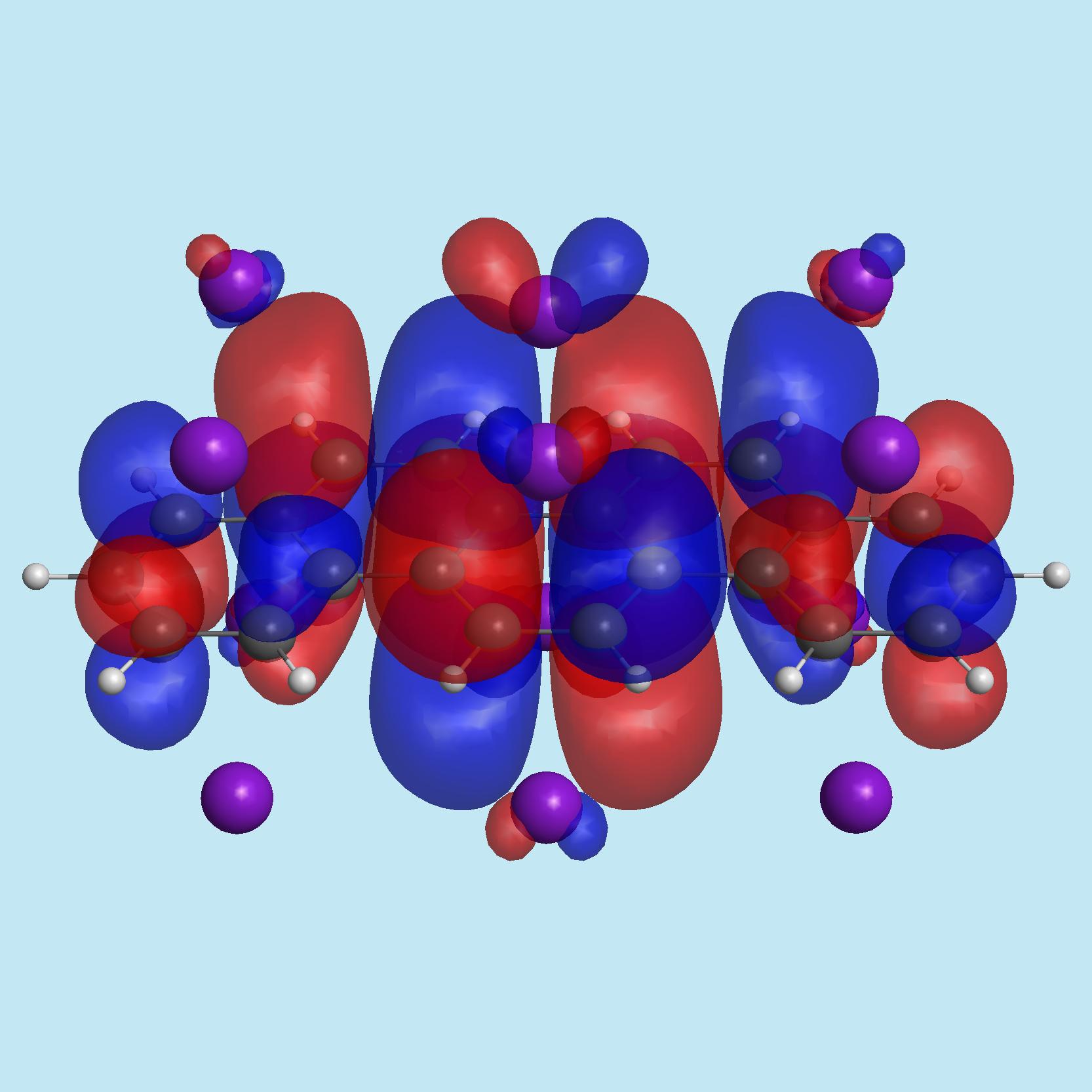}&
\includegraphics[width=0.49\columnwidth]{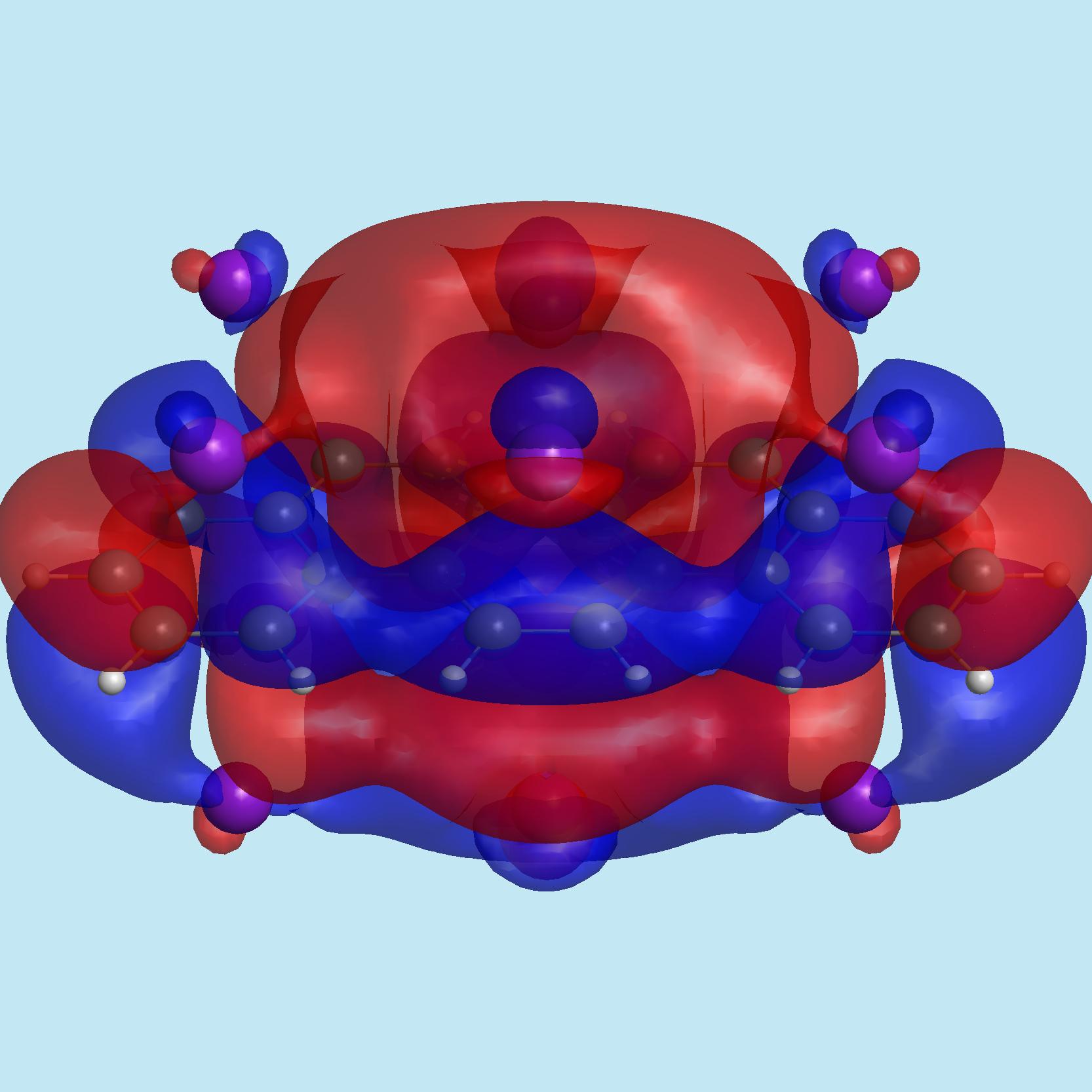}\\
\end{array}$
\caption{(Color online) Most relevant Natural Orbitals obtained
for the K$_{12}$picene$^{9+}$ cation, that is, a cluster
populated by 362 core electrons and three valence electrons\cite{core}.
Orbital occupations are 1.941 (left orbital) and 0.992 (right orbital).
Higher energy orbitals show tiny occupations.
The connection between electronic occupation
of crystalline K$_3$picene and the number of electrons in clusters
is discussed within the text (see subsection IIIA). 
}
\label{fig:K12picene}
\end{figure}

In one case we have added chlorine atoms to the cluster
to avoid changing the total number of electrons of the system after
geometrical optimization. One electron is transferred from potassium
atoms to each chlorine atom that is added to the cluster. Therefore,
a Cl$_{12}$K$_{18}$picene$_2$ cluster has six valence electrons ready
to populate picene LUMOs. However, DFT results show that two of these
electrons remain on K atoms again demonstrating that the electron
transfer from K to picene is not complete. Following a similar reasoning
as used above for binary clusters, a complete charge consistency
would imply some reduction of the number of chlorine atoms. 
This implies further geometry optimization and still lower symmetry
around picene molecules, that is, strong reasons for not insisting
along this line.

The electronic structure of clusters has been obtained using
US GAMESS\cite{GAMESS,GAMESSbis} program that is a complete quantum chemistry package
that allows both Density Functional Theory (DFT) calculations and precise
Configuration Interaction (CI) multireference calculations.
When performing mean-field calculations,
exchange and correlation are approximately included
by the B3LYP functional\cite{b3lyp,b3lypbis,b3lypbisbis}.
This is an hybrid GGA (Generalized Gradient Approximation)
functional combining five functionals, in particular,
Hartree-Fock (HF) exchange.
A large valence triple zeta plus polarization on all atoms (6-311G**)
gaussian basis has been chosen for H, C, and K atoms\cite{base1,base2}.
All graphical results have been obtained using the WXMACMOLPLT
package\cite{macmolplt}.

\section{Quantum chemistry results}

\subsection{Picene molecule in a potassium environment}

This is the smallest K-picene system we have studied and the only one allowing
state-of-the-art CI analysis. Based on the results of this case, we suggest
that multireference calculations can be partially mimic at the DFT level
increasing the spin multiplicity of the electronic state.
In this way electrons can visit
molecular orbitals that are not occupied in the singlet state.
Sometimes even the total electronic energy decreases with increasing
spin multiplicity.
We do not give any significance to this fact since most probably
a CI analysis would change the conclusions about the correct spin of the
ground state. 

To learn about the effect of K intercalation in picene crystal, one
picene molecule is surrounded by twelve K atoms as it happens in the crystalline
structure. The positions of K atoms are forced to conserve the
original C$_{2v}$ symmetry of the isolated polycyclic aromatic molecule
just to make easier both computation and analysis.
The precise geometry is obtained by minimization of the B3LYP total energy
of the neutral cluster (the resulting cluster is represented
in the upper left panel of Fig. \ref{fig:clusters}). The final geometry
does not differ appreciably from the crystalline one\cite{xxx}.
In fact,
the mean distance of potassium atoms to the plane of the picene molecule
is 2.91 \AA, being deviations from this value less than 0.3 \AA.
Similar values hold for the other clusters shown in
Fig. \ref{fig:clusters}\cite{geo}.
After geometry optimization of the neutral cluster,
nine electrons are removed from the cluster limiting
to three the maximum number of electrons that potassium can
transfer to the picene molecule.
As has been said before, electrons removed from the cluster belong
to neighboring cells that are not explicitly included in the study.
Geometry is hold fixed as a further optimization would
artificially increase K-picene distances.

\begin{figure}
$\begin{array}{cc}
\includegraphics[width=0.49\columnwidth]{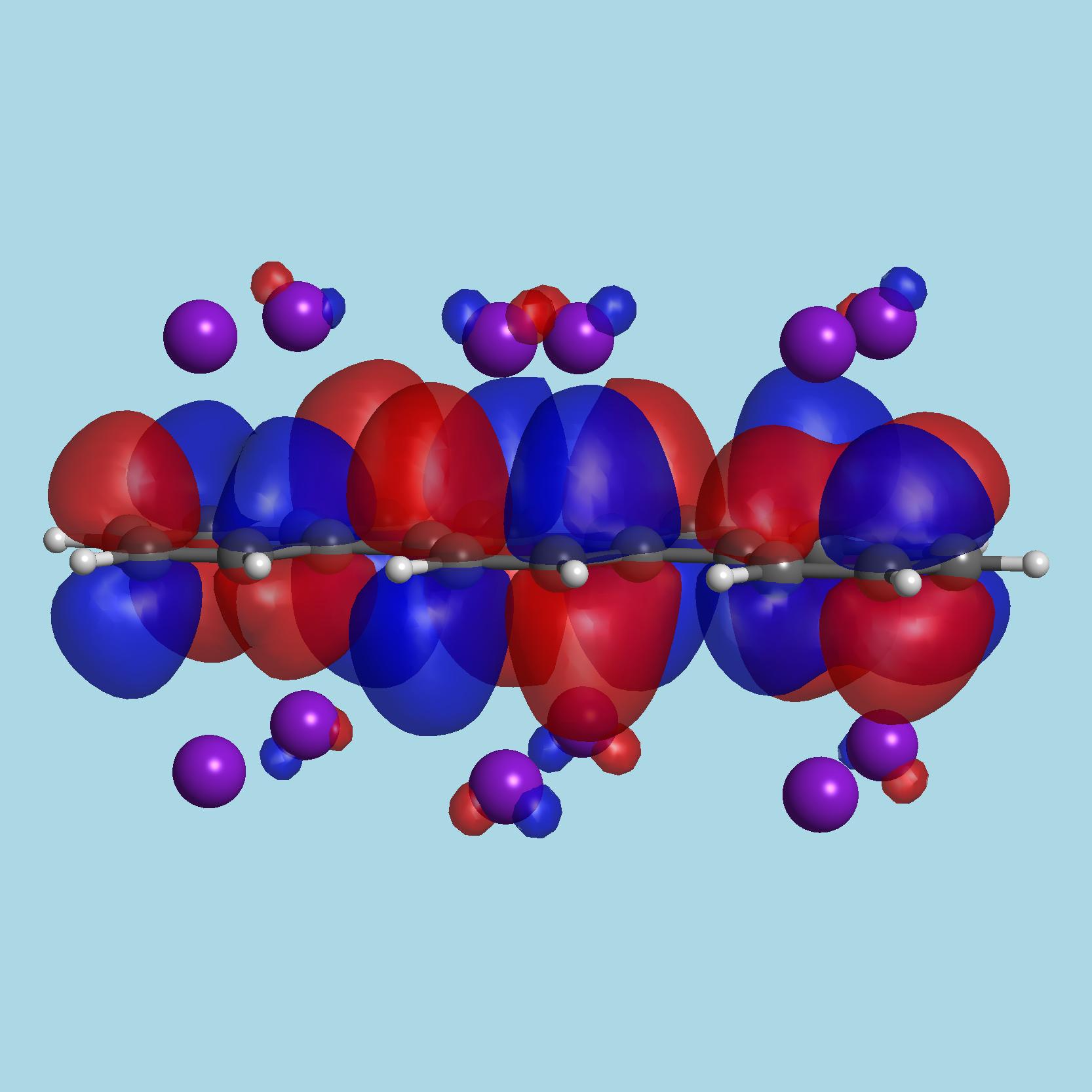}&
\includegraphics[width=0.49\columnwidth]{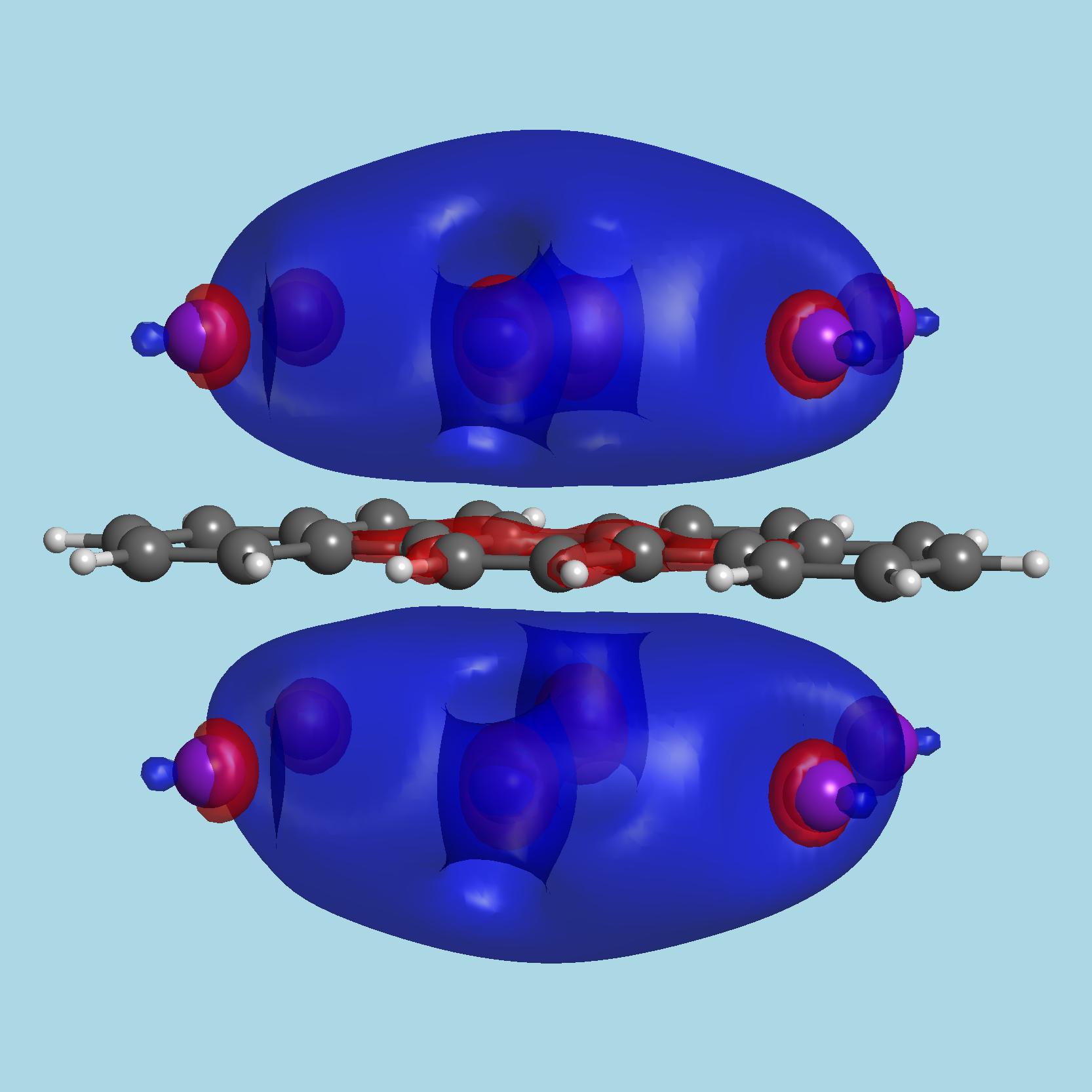}\\
\includegraphics[width=0.49\columnwidth]{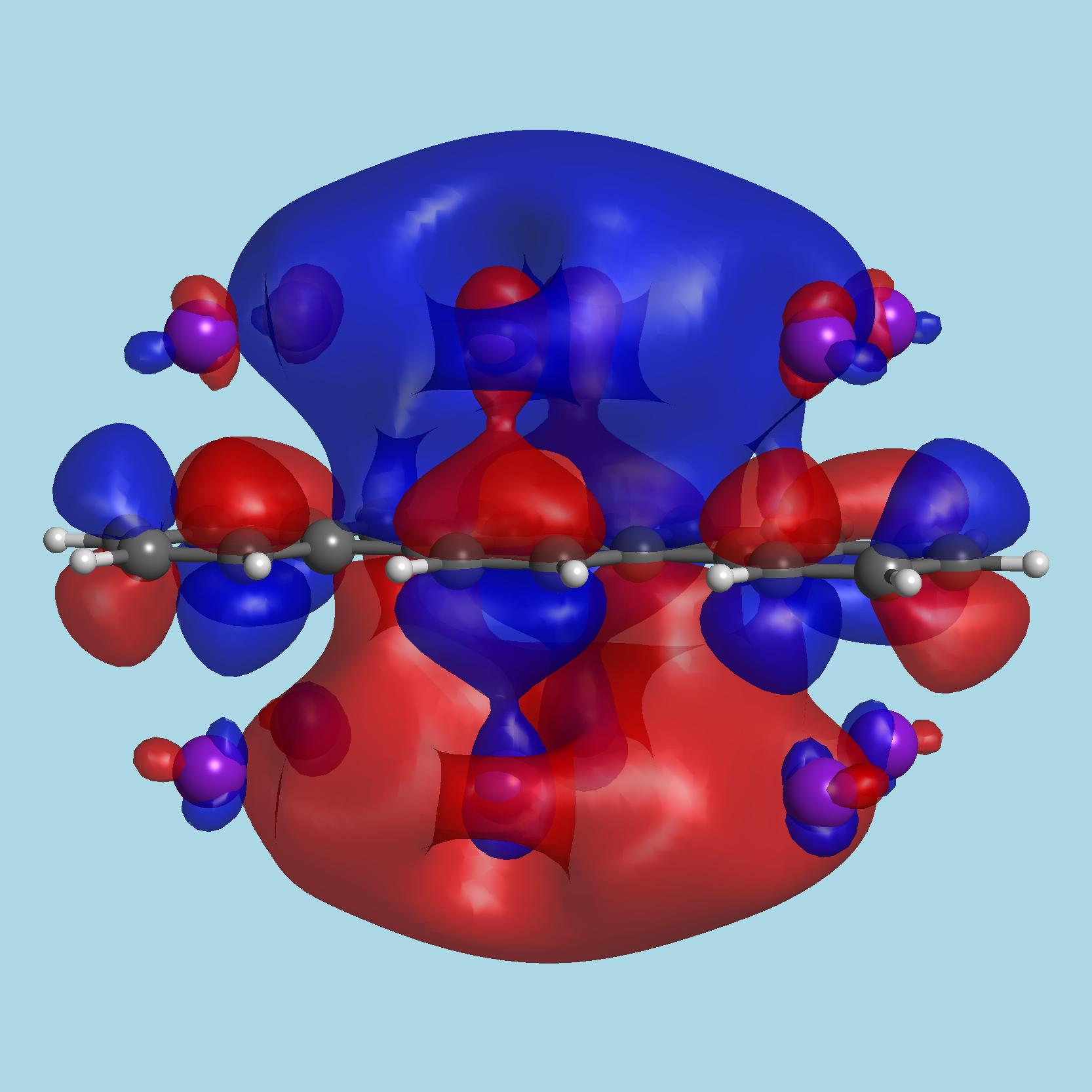}&
\includegraphics[width=0.49\columnwidth]{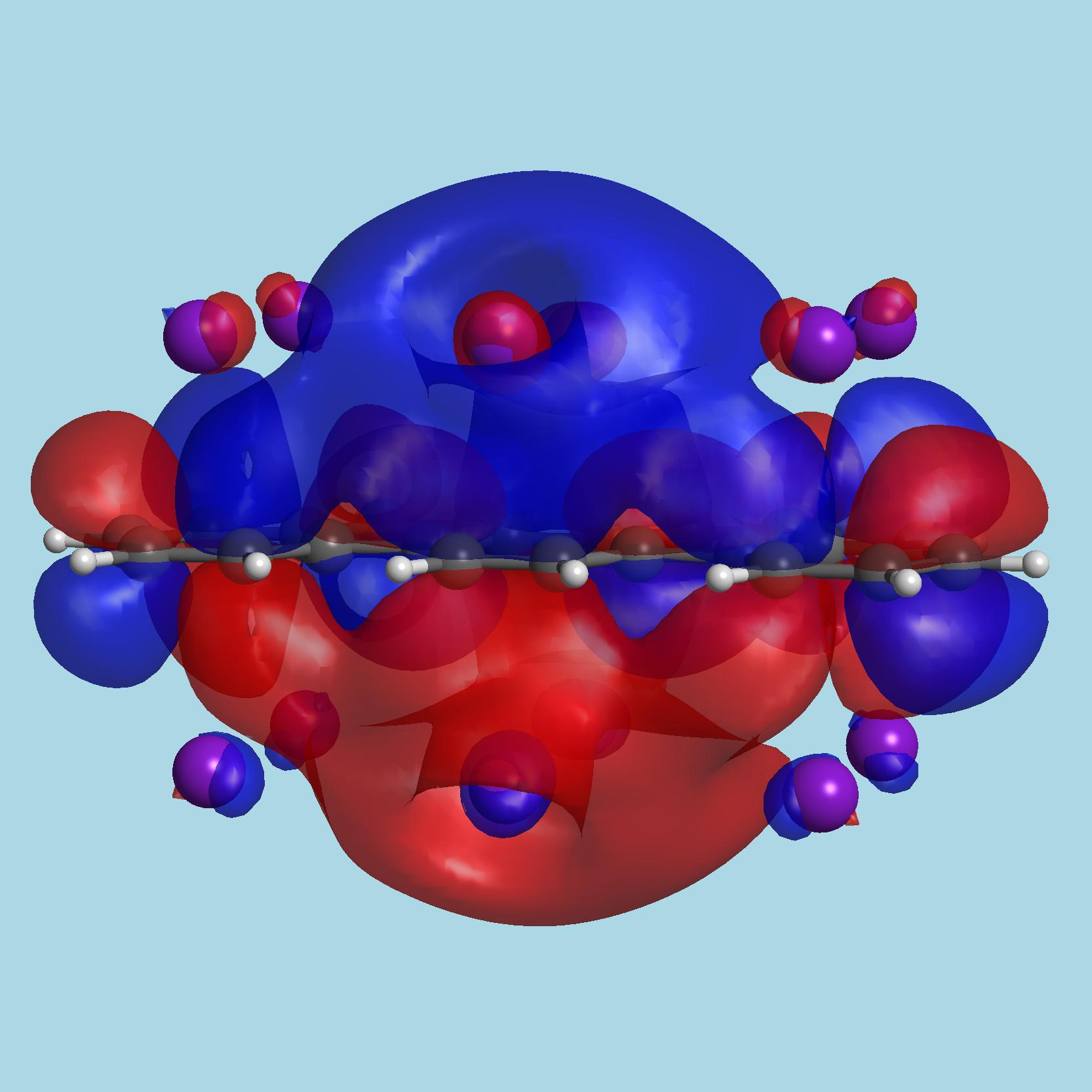}\\
\includegraphics[width=0.49\columnwidth]{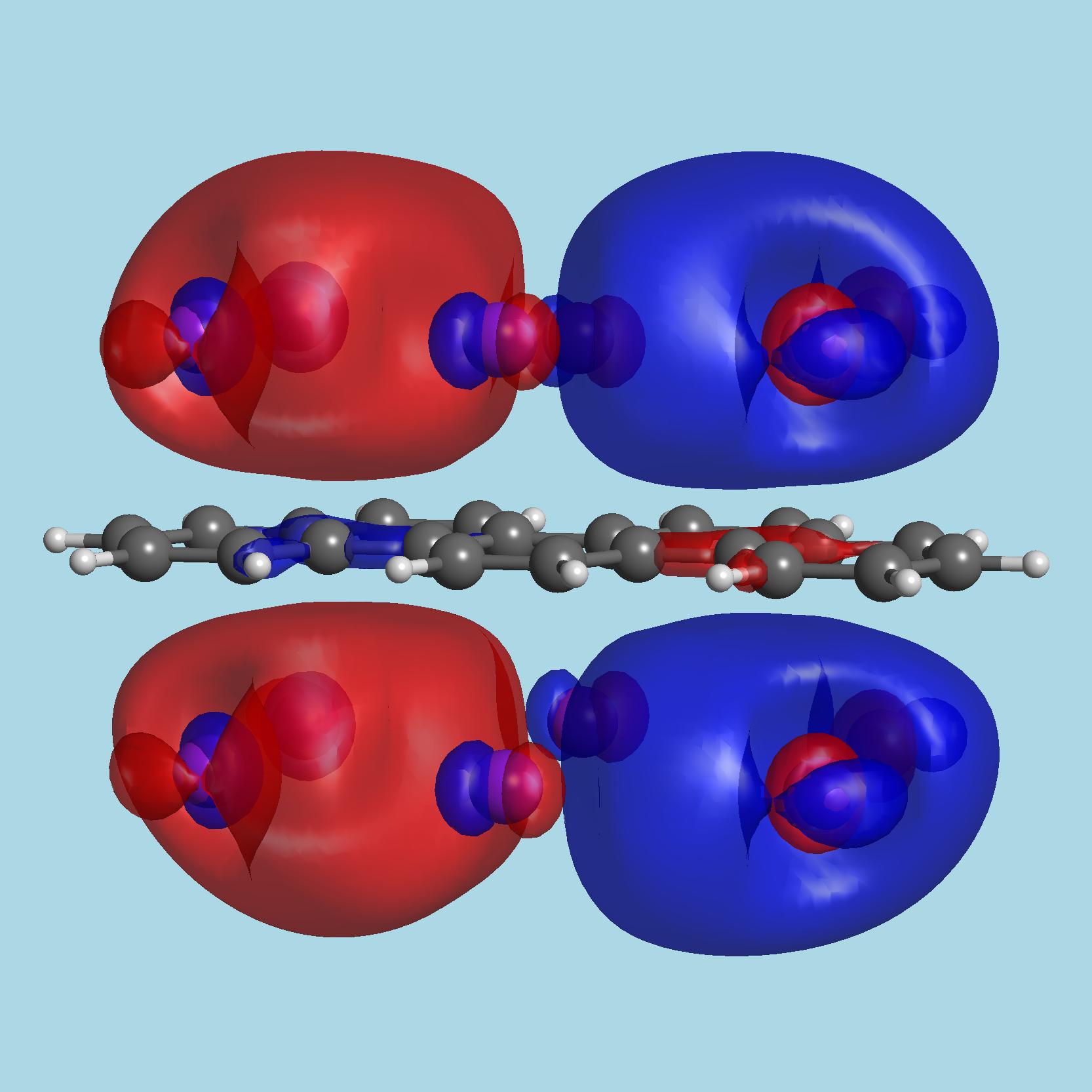}&
\includegraphics[width=0.49\columnwidth]{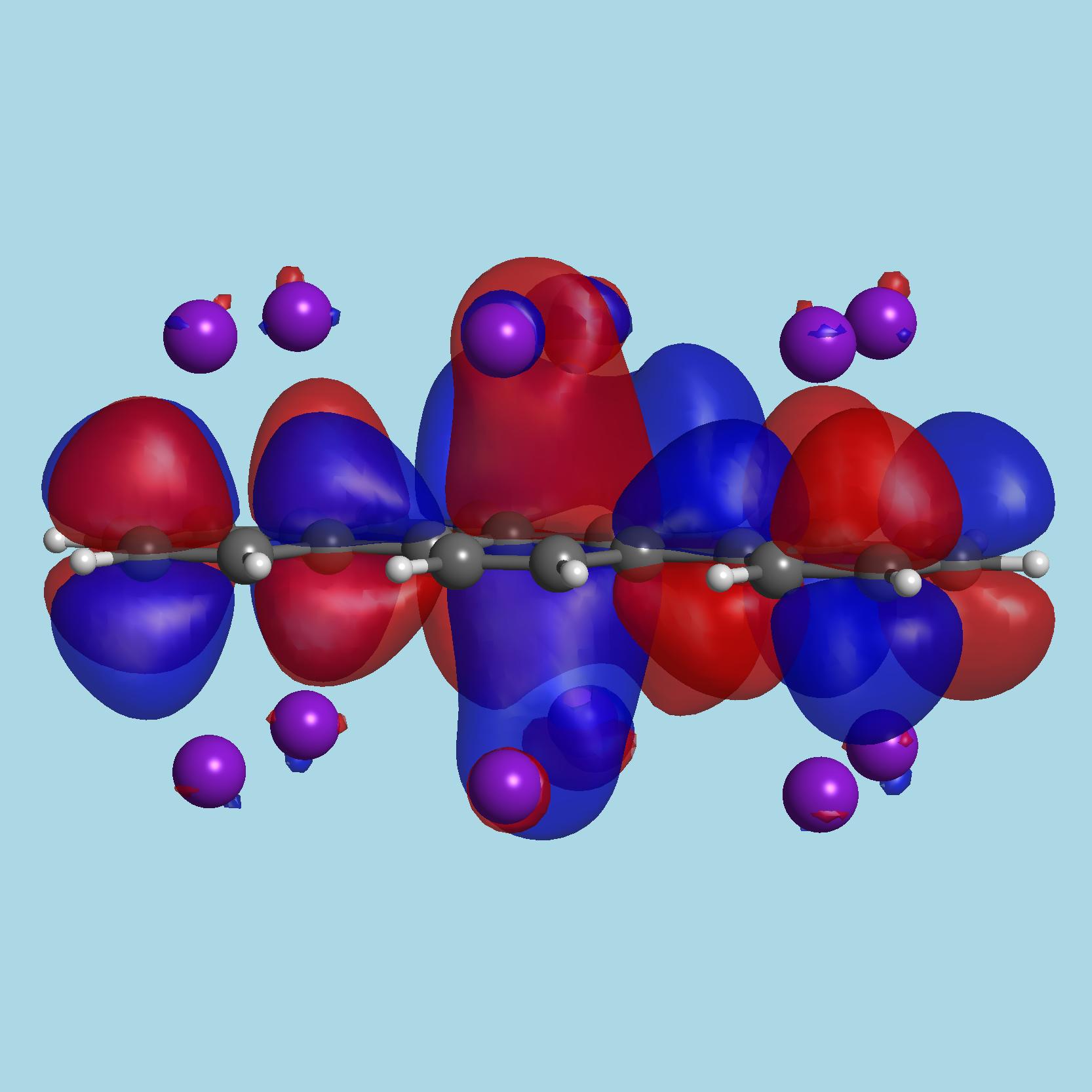}\\
\end{array}$
\caption{(Color online) Molecular orbitals corresponding to valence
electrons as obtained for
the ground state (a spin triplet) of the K$_{12}$picene$^{8+}$ cluster.
Following the numeration scheme stated in the caption of Fig. \ref{fig:clusters},
the first orbital is doubly occupied,
the next two are singly occupied and the last three are empty.
}
\label{fig:K12piceneBIS}
\end{figure}

Our B3LYP calculations for the K$_{12}$picene$^{9+}$ cation
indicate that while two electrons occupy the LUMO
of isolated picene molecule, the third one remains in a symmetric combination
of $4s$ K orbitals (more precisely, the bonding combination of all twelve $4s$
orbitals with almost equal coefficients that is shown in the upper right
panel of Fig. \ref{fig:K12piceneBIS} for the somewhat less
charged K$_{12}$picene$^{8+}$ cation).
Nevertheless, picene LUMO+1 and LUMO+2 are
close in energy suggesting that they will play an important role
if the one-configuration limitation is avoided.
In fact, some tweaking of minor computational details like
geometry optimization or not of the ion, or keeping or not the
symmetry group shows that the third valence
electron can either occupy K orbitals or flow to some
combination of picene LUMO+1 and LUMO+2.
Consequently, a robust description of the electronic state
requires at least a two-configuration wavefunction participated
by both K and picene orbitals.
Here, DFT analysis was followed by a precise MCSCF (Multi-Configurational
Self Consistent Field) wavefunction study
in which sixteen orbitals describing the three valence electrons\cite{core}
were allowed to mix. That means 468 (480) determinants having space
symmetry A$_1$ (B$_2$) for $S_{\rm z}=\frac{1}{2}$.
Next, we will discuss the results obtained for the more interesting B$_2$
symmetry although the lowest energy obtained for this symmetry
is somewhat higher (0.31 eV) than the one obtained for the A$_1$ wavefunction.
We choose B$_2$ solution for symmetry reasons since A$_1$ K orbitals do not couple
to $\pi$ electrons. The occupation of an A$_1$ molecular orbital
(upper right panel of Fig. \ref{fig:K12piceneBIS}) would lead to a much larger charge on
potassium atoms, i.e., to an almost pure 4s K band showing half filling.
May be this could provide an alternative
explanation for metallic conductivity but we are strongly inclined to think that an
ionic scenario showing partial charge transfers from K to picene is the most adequate.

Fig. \ref{fig:K12picene} shows the natural orbitals that describe the three
valence electrons originating from K intercalation.
The first one (left panel) is very similar to the LUMO
of isolated picene molecule and its occupation is 1.941. It is antisymmetric
with respect to a plane cutting picene molecule (and whole cluster)
in two identical halves. On the other hand,
the second one (right panel) is a mixture of picene  LUMO+1 and LUMO+2
(both of B$_2$ symmetry) and $4s$ potassium orbitals. It is symmetric relative to the
above mentioned plane and its occupation is 0.992. Its hybrid character
indicates that the third electron does not flow completely to picene orbitals but
remains partially on K atoms. Since good quality CI computation is not
possible for larger clusters, we will explore now
an easier way to partially recover these results.
Firstly, we argue that some additional electrons should be
included in the cluster because our MCSCF calculation proves that
only about two and a half electrons are transferred to the picene
molecule and, consequently, about one half electron
remains on every group formed by three K atoms. Therefore, the
total count of valence electrons is $2.5+4 \times 0.5=4.5$ in this case,
that is, one and a half more electrons than considered in the previous
multi-configurational analysis.
As subsequent calculations at a DFT level evidence, the addition of
just one extra valence electron to the system suffices to
end with a nearly consistent charge both on picene and K
groups. Let us remark here that although this is our preferred
choice, main conclusions are the same for three, four or five
valence electrons. 
Secondly, larger spin degeneracies that partially populate higher molecular orbitals
are considered. In this case, a spin triplet provides
the best total electronic energy from a single-determinantal point of view. The
calculated MOs are shown in Fig. \ref{fig:K12piceneBIS}. The first one
is again picene LUMO, the second one is mainly $4s$ K while third (singly occupied)
and fourth (empty)
show the characteristics we are looking for in our study. Both are antisymmetric
relative to the plane containing the picene molecule and both show an important
hybridization between picene and potassium orbitals.
As a matter of fact,
the empty fourth MO is very similar to the occupied one
obtained in the more precise MCSCF description (see Fig. \ref{fig:K12picene}).
This orbital becomes singly occupied for a spin quintuplet but the
energy price paid for the excitation of one of the electrons
on the LUMO is important (Table I gives the details).  
A rough electron counting based exclusively in the pictures gives 2.5 electrons
on picene and 1.5 electrons on K atoms which is satisfactorily consistent
with our assumption for the distribution of the valence charge.
 
\begin{figure}
$\begin{array}{ccc}
\includegraphics[width=0.32\columnwidth]{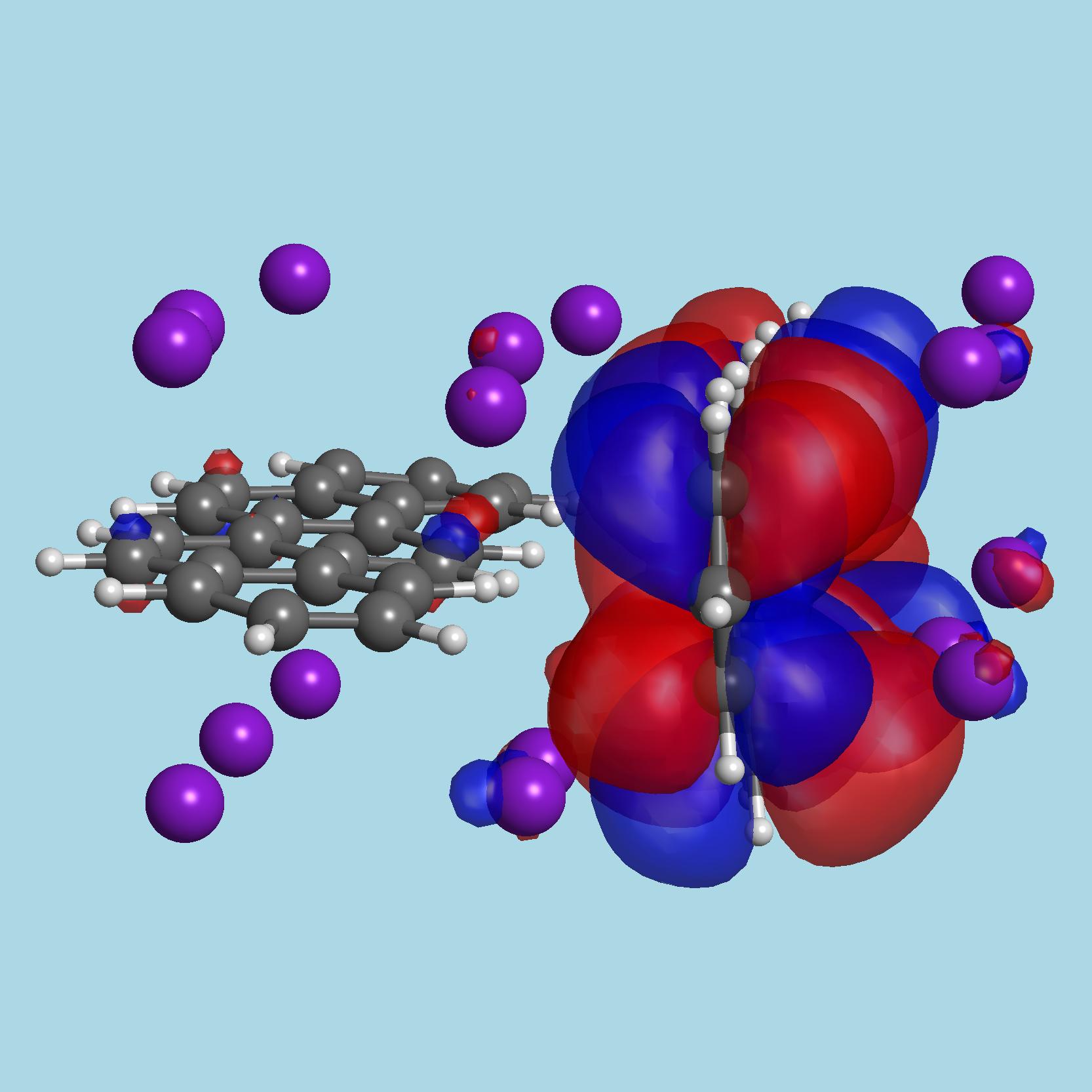}&
\includegraphics[width=0.32\columnwidth]{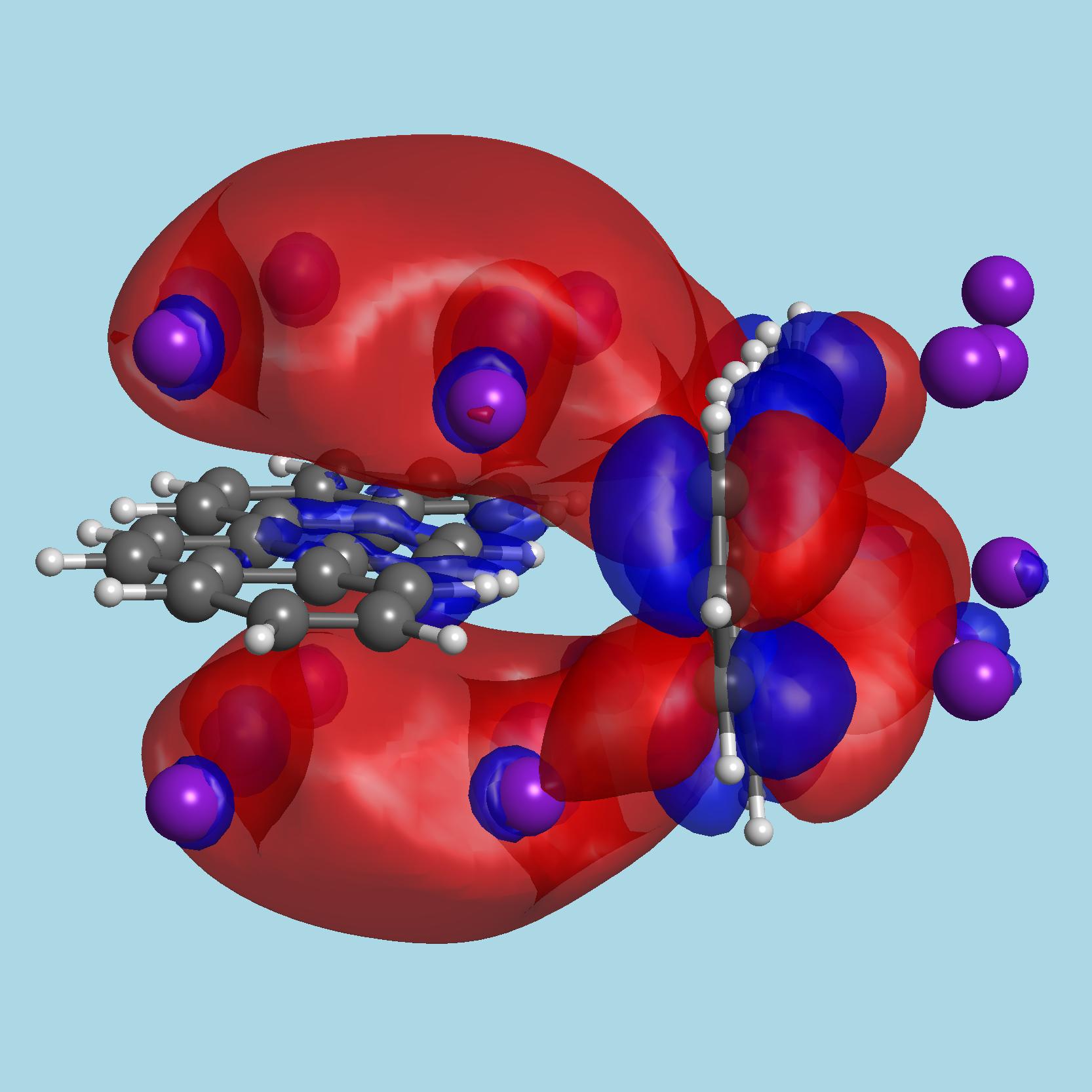}&
\includegraphics[width=0.32\columnwidth]{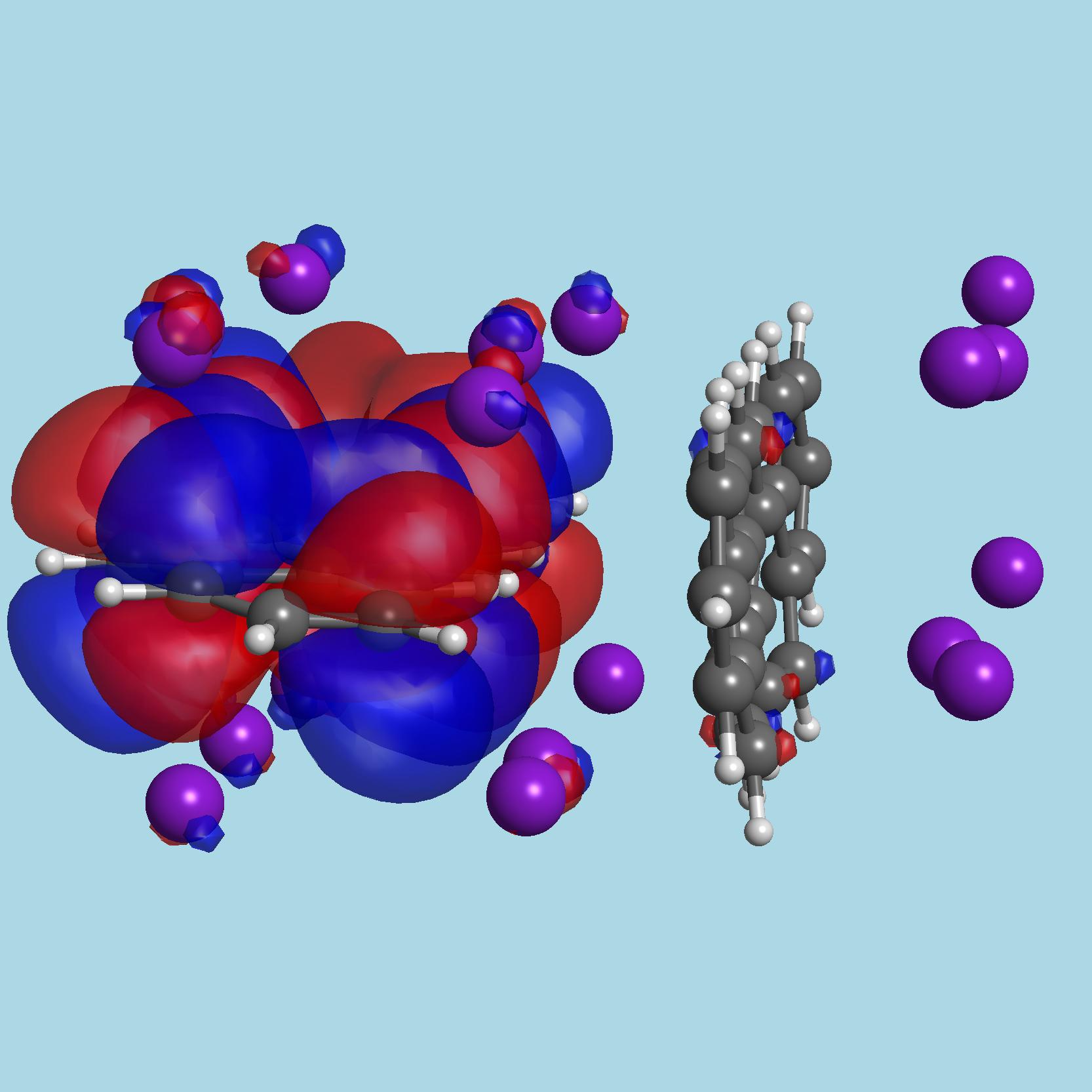}\\
\includegraphics[width=0.32\columnwidth]{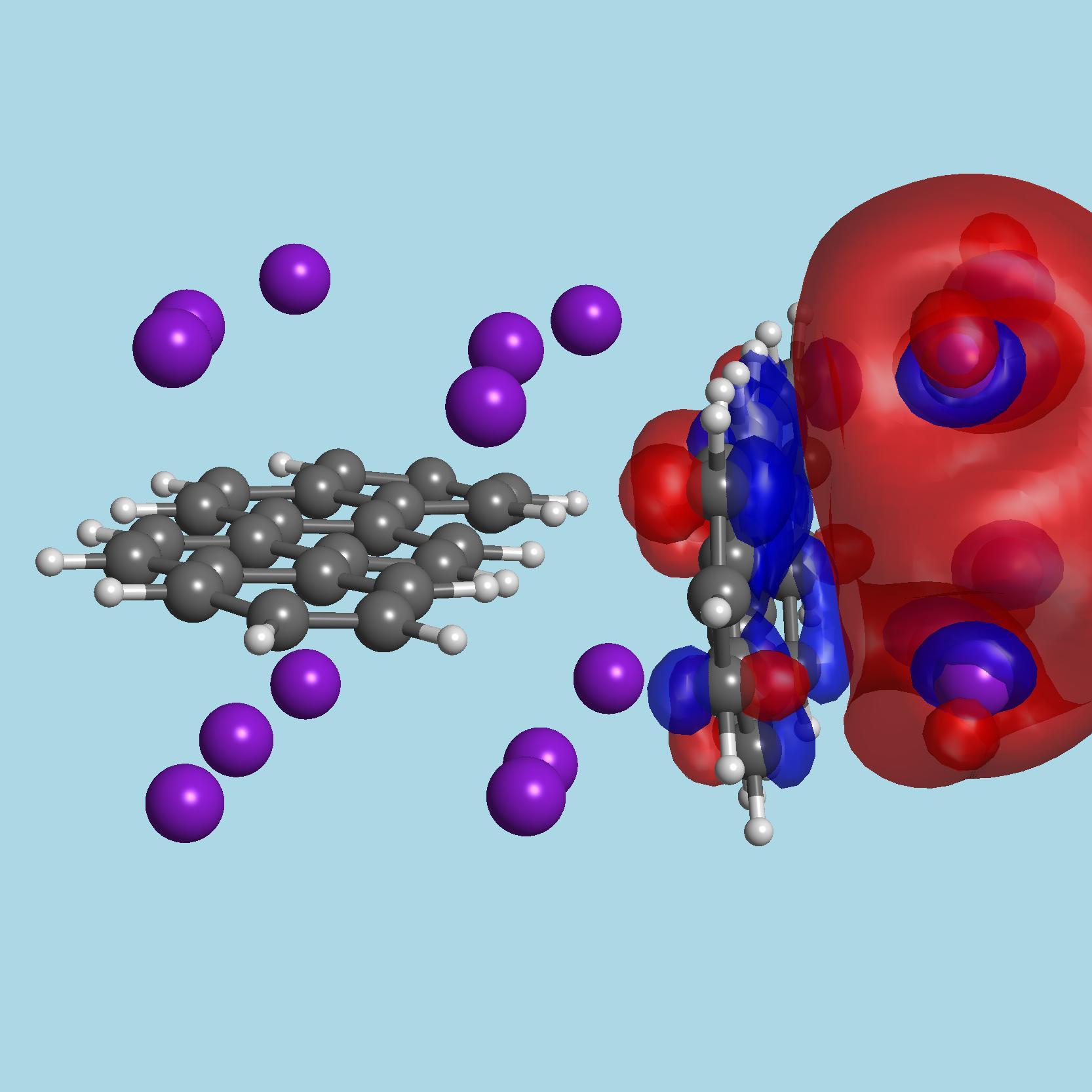}&
\includegraphics[width=0.32\columnwidth]{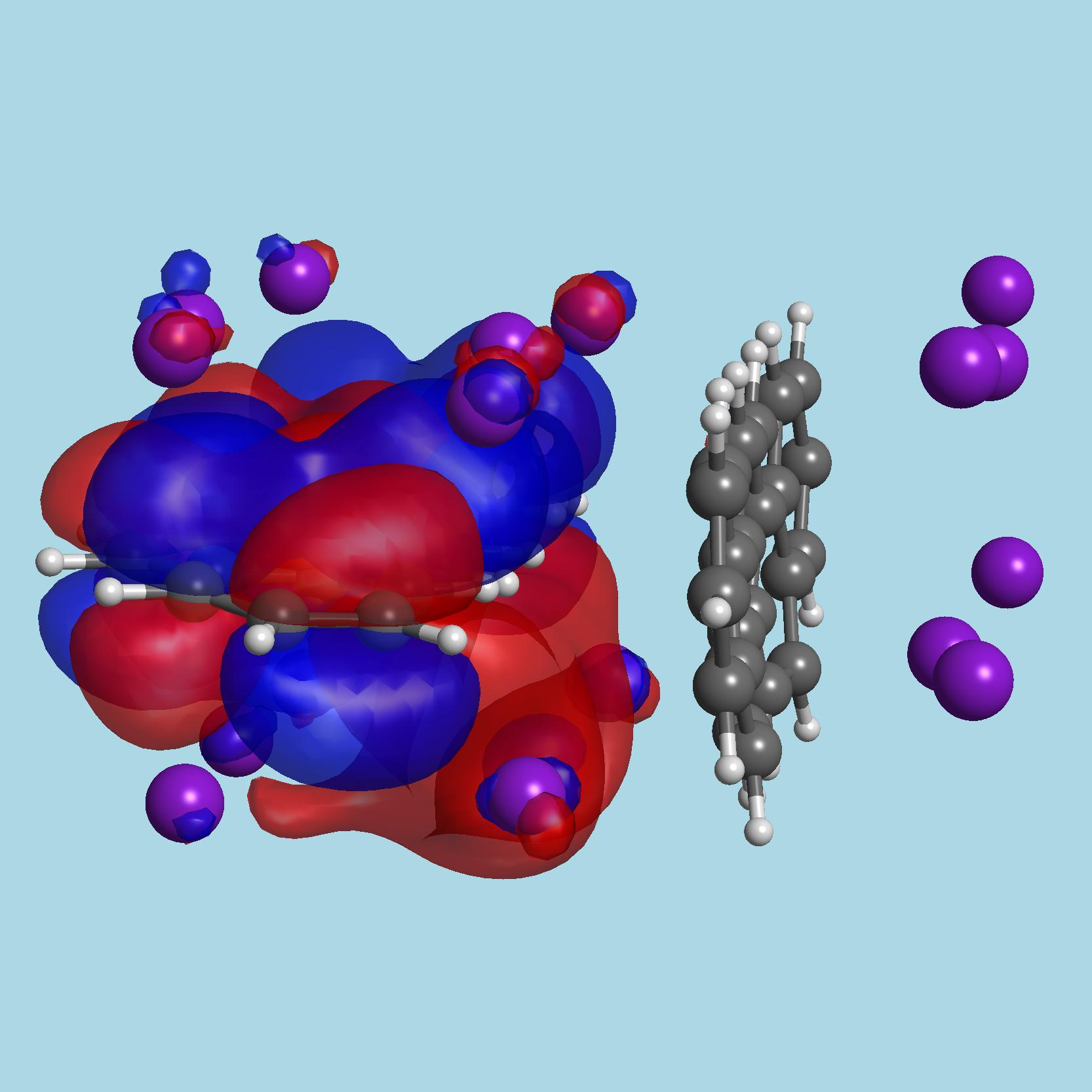}&
\includegraphics[width=0.32\columnwidth]{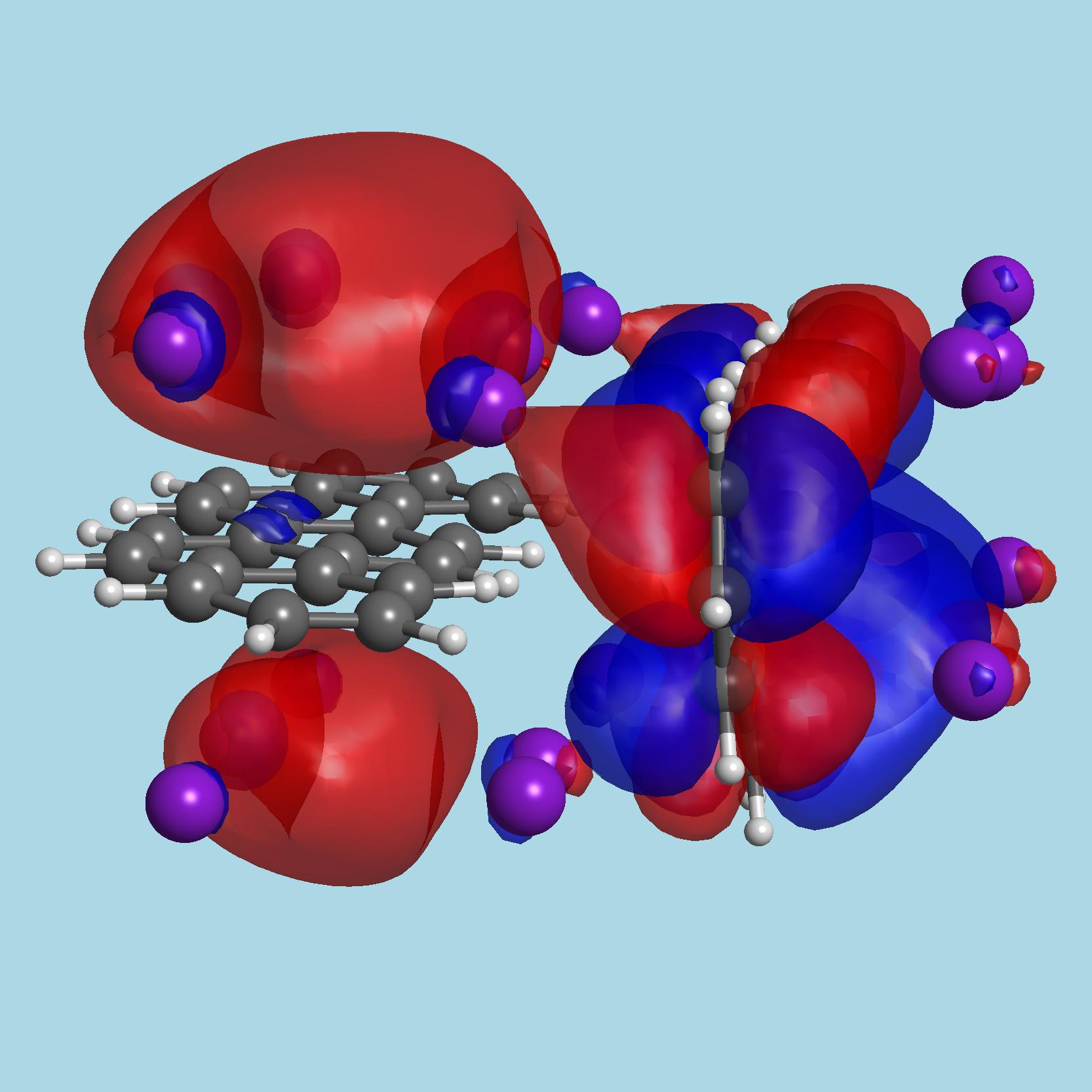}\\
\includegraphics[width=0.32\columnwidth]{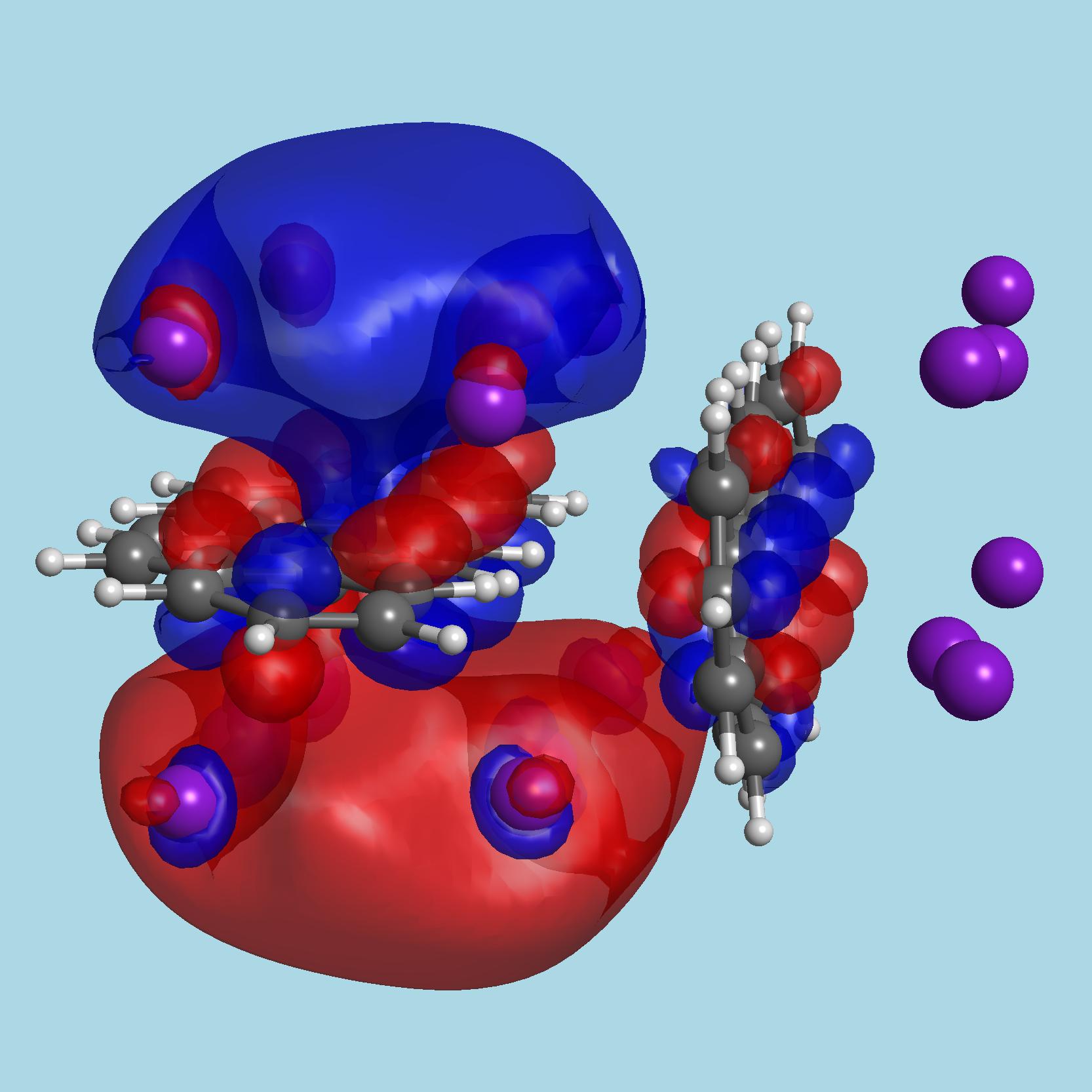}&
\includegraphics[width=0.32\columnwidth]{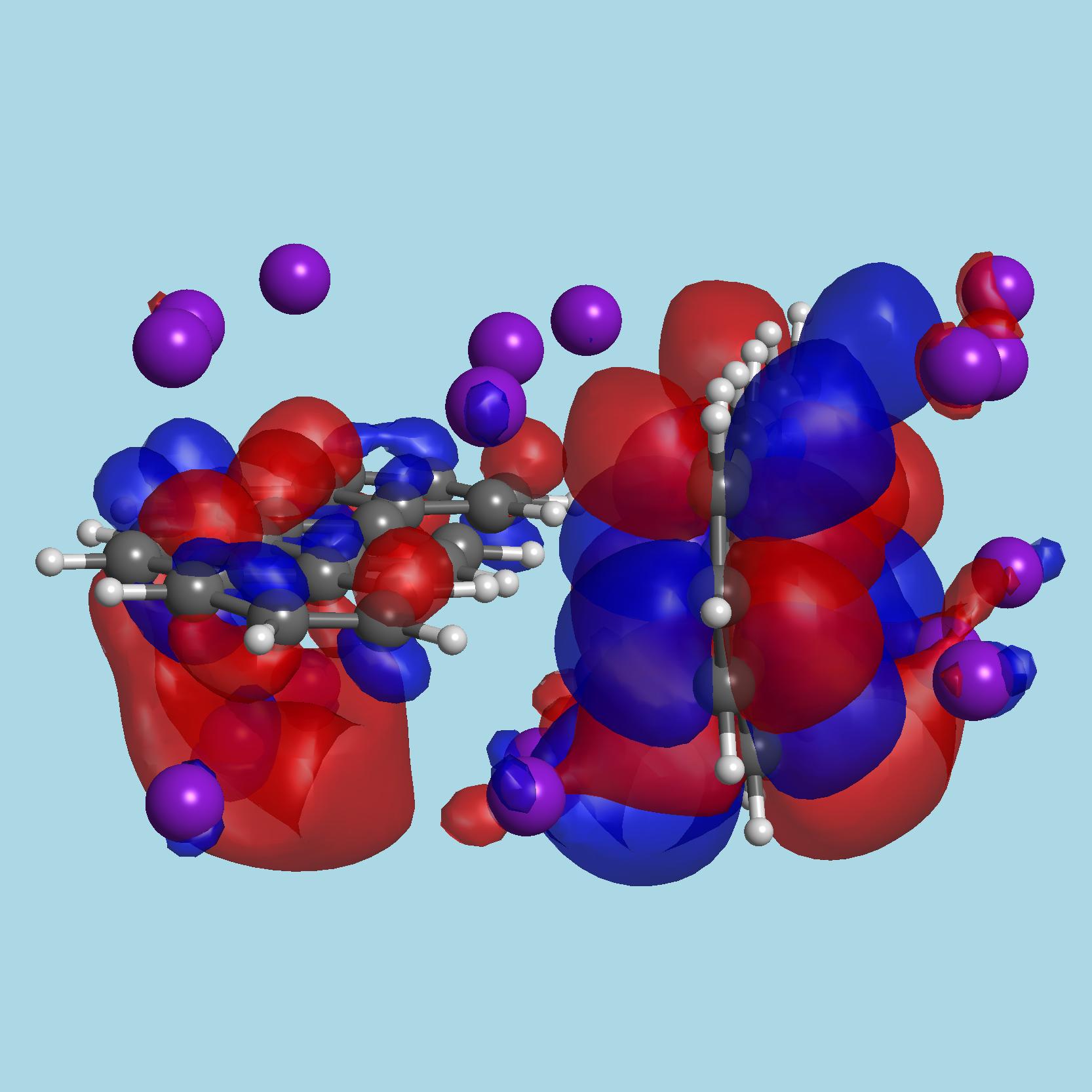}&
\includegraphics[width=0.32\columnwidth]{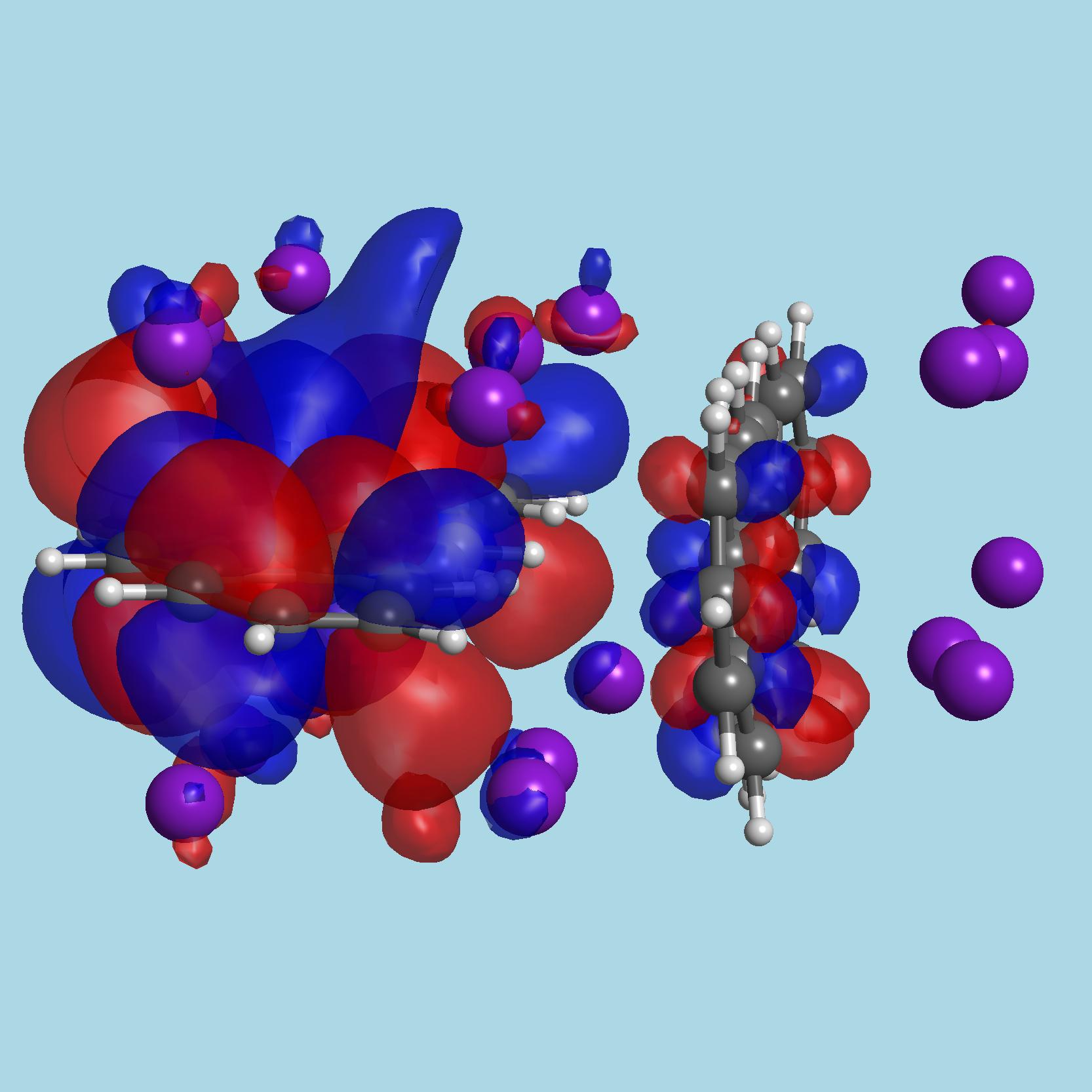}\\
\end{array}$
\caption{(Color online) Molecular orbitals of valence electrons
obtained for the ground state (a spin triplet) of the
K$_{18}$picene$_2$$^{10+}$ cluster.
Following the numeration scheme stated in the caption of Fig. \ref{fig:clusters},
the first three orbitals are doubly occupied,
the next two are singly occupied and the rest are empty.
}
\label{fig:K18picene2}
\end{figure}

\subsection{K$_{18}$picene$_2^{10+}$ cation}

In this subsection, the results for two picene molecules in an edge to
face conformation surrounded by
eighteen potassium atoms are presented (upper right panel of
Fig. \ref{fig:clusters} shows the cluster). The geometry of the cluster
has been optimized using DFT for the neutral system.
Afterward, the electronic structure of a
cluster with less electrons in the outer shell has been studied.
Based in our experience, we assume that half an electron
remains on every K group formed by three atoms. That means
a total number of three electrons on K atoms
that added to the
five electrons transferred to picene molecules gives a total number
of eight valence electrons in the cluster. Therefore, ten electrons are
removed from the neutral cluster to end with a $10+$ cation.
Several spin multiplets have been obtained for this system, being
a spin triplet its ground state. Molecular orbital energies corresponding
to valence electrons are about 3 eV above the highest core electron
molecular orbital energy.
Fig. \ref{fig:K18picene2} shows the first nine MOs in the
valence band region. The first and the third correspond to LUMO
of neutral picene while the rest show an important K-picene
hybridization. It is easy to see that the number of electrons on K atoms
is fully consistent with our initial assumption of just three electrons
on K atoms. Table I gives a detailed account of the low energy spin
excitations of this system. The role played by electronic excitations
will be discussed later.  

\begin{figure}
$\begin{array}{ccc}
\includegraphics[width=0.32\columnwidth]{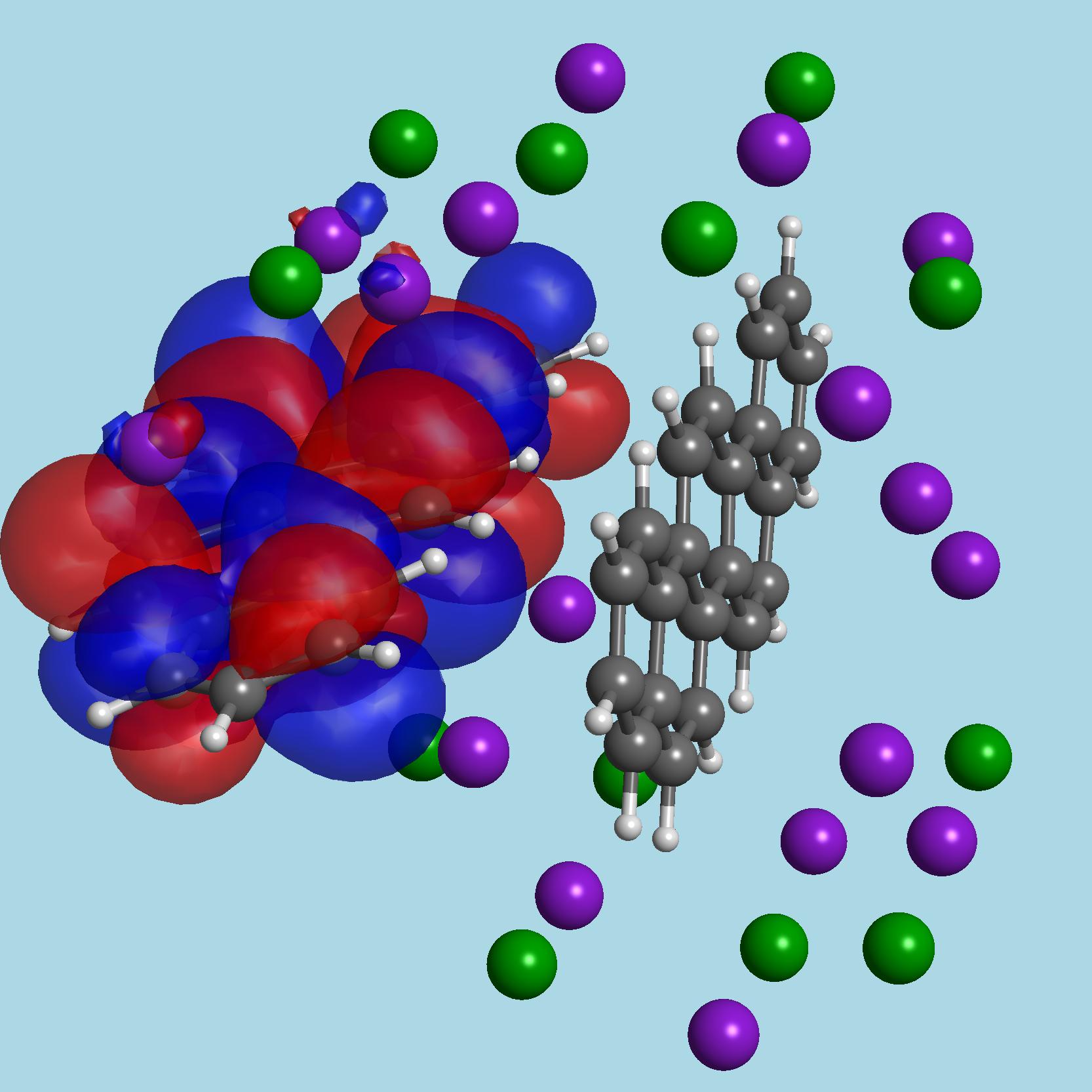}&
\includegraphics[width=0.32\columnwidth]{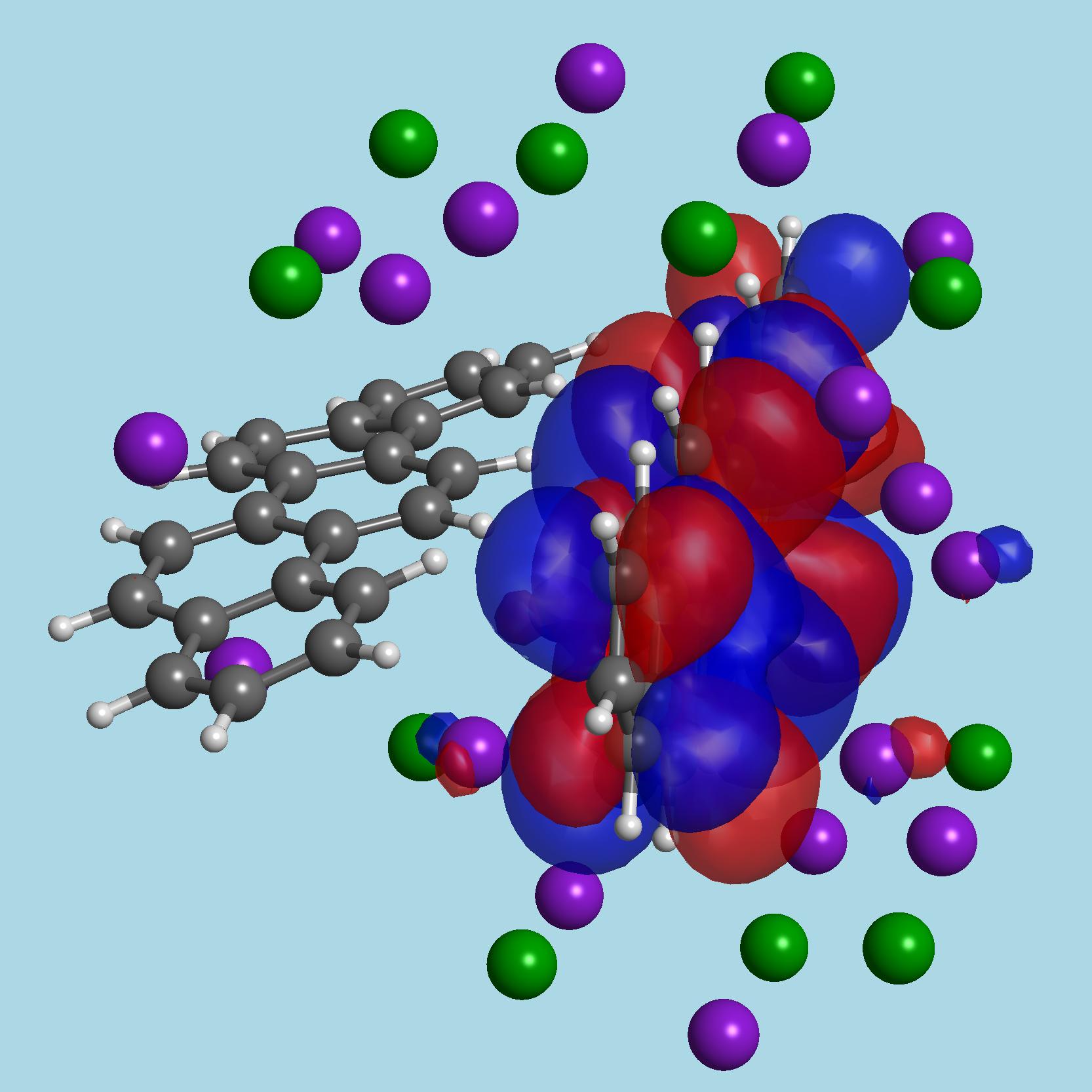}&
\includegraphics[width=0.32\columnwidth]{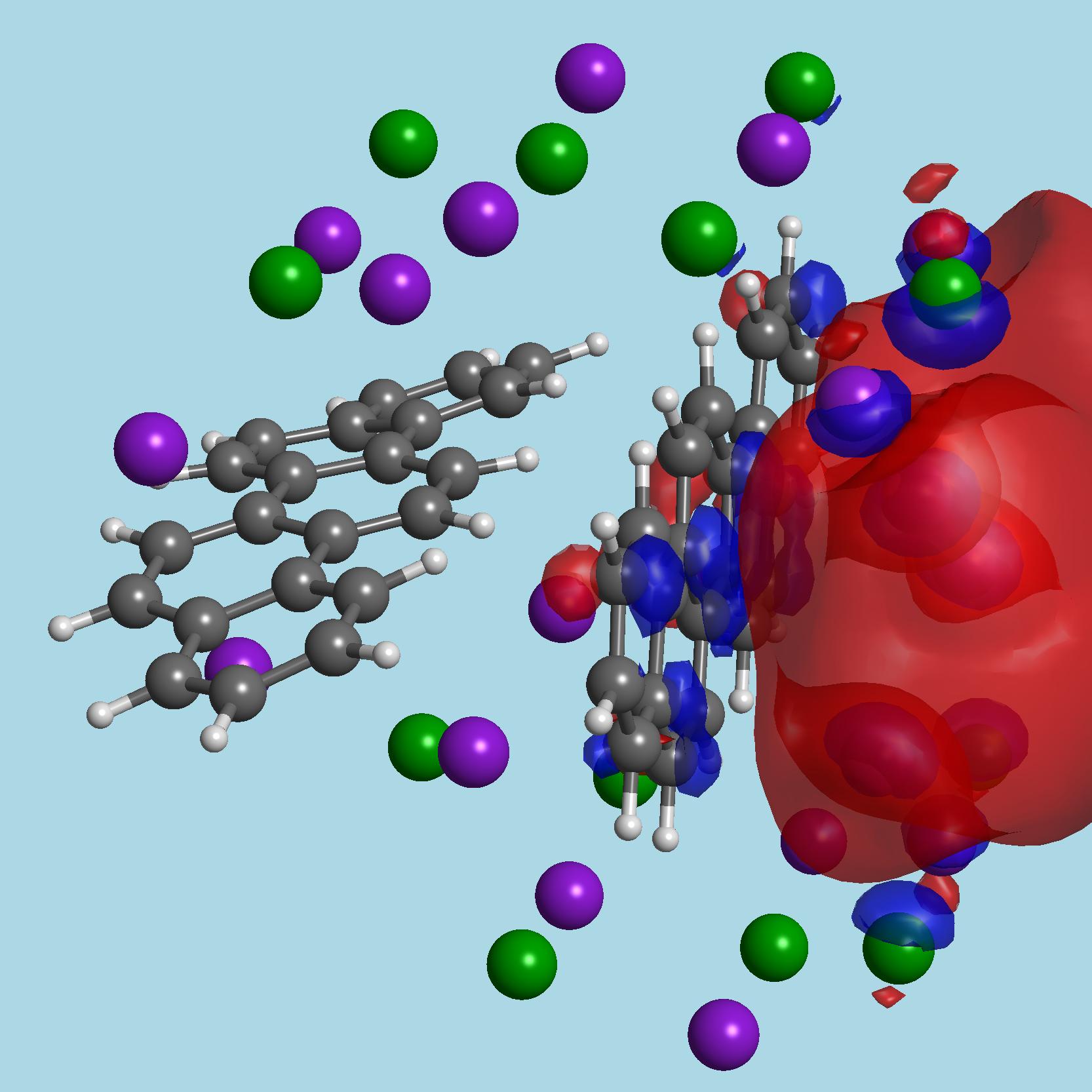}\\
\includegraphics[width=0.32\columnwidth]{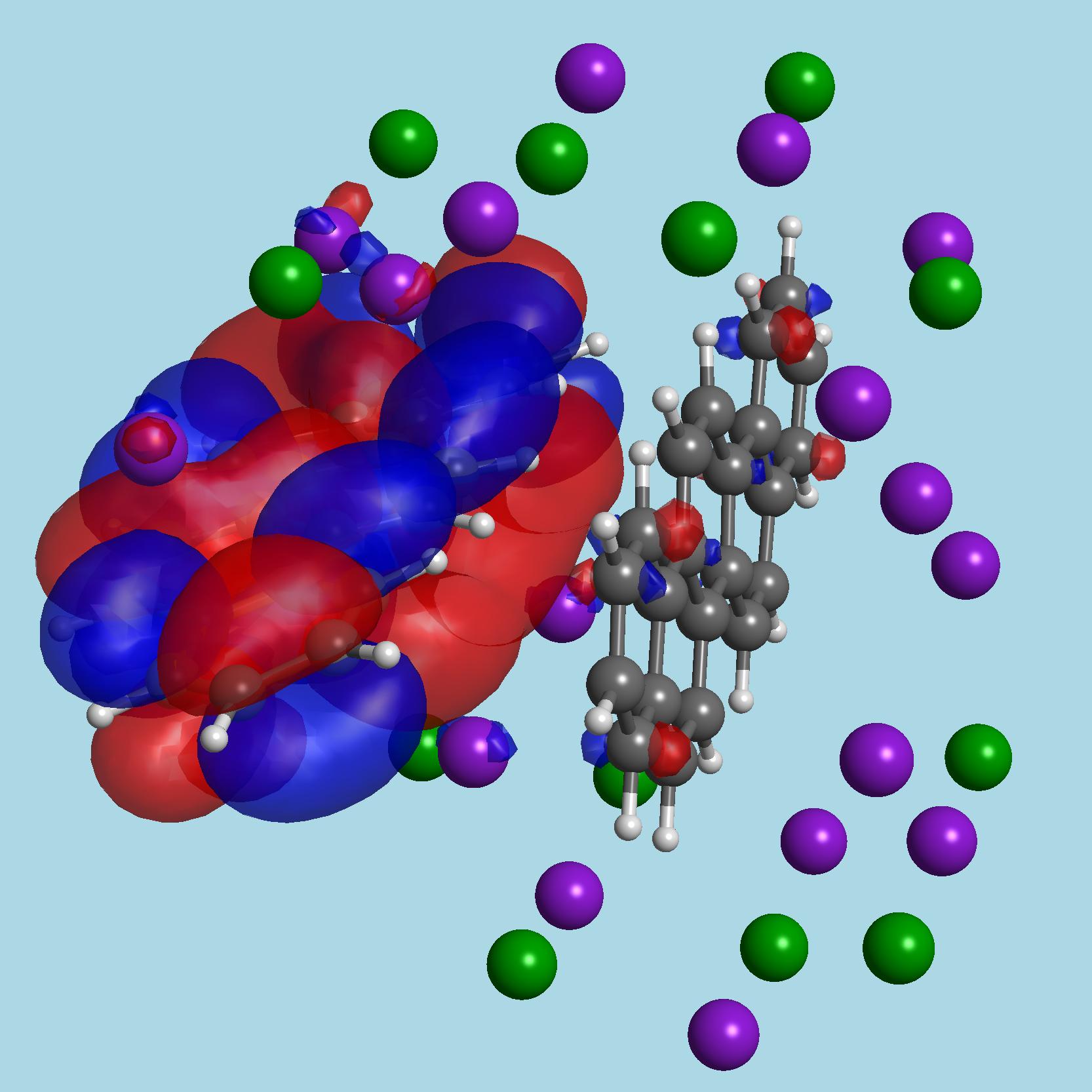}&
\includegraphics[width=0.32\columnwidth]{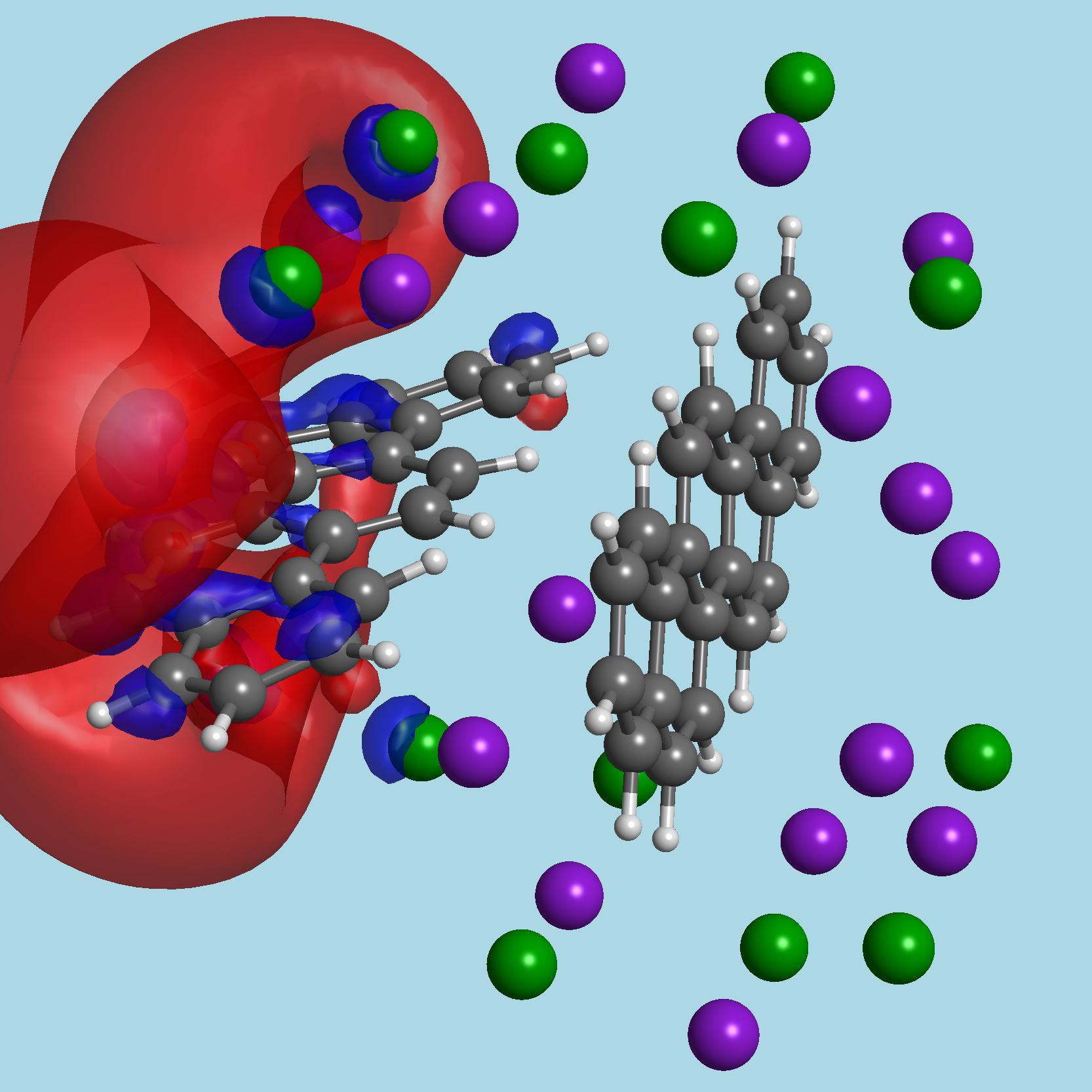}&
\includegraphics[width=0.32\columnwidth]{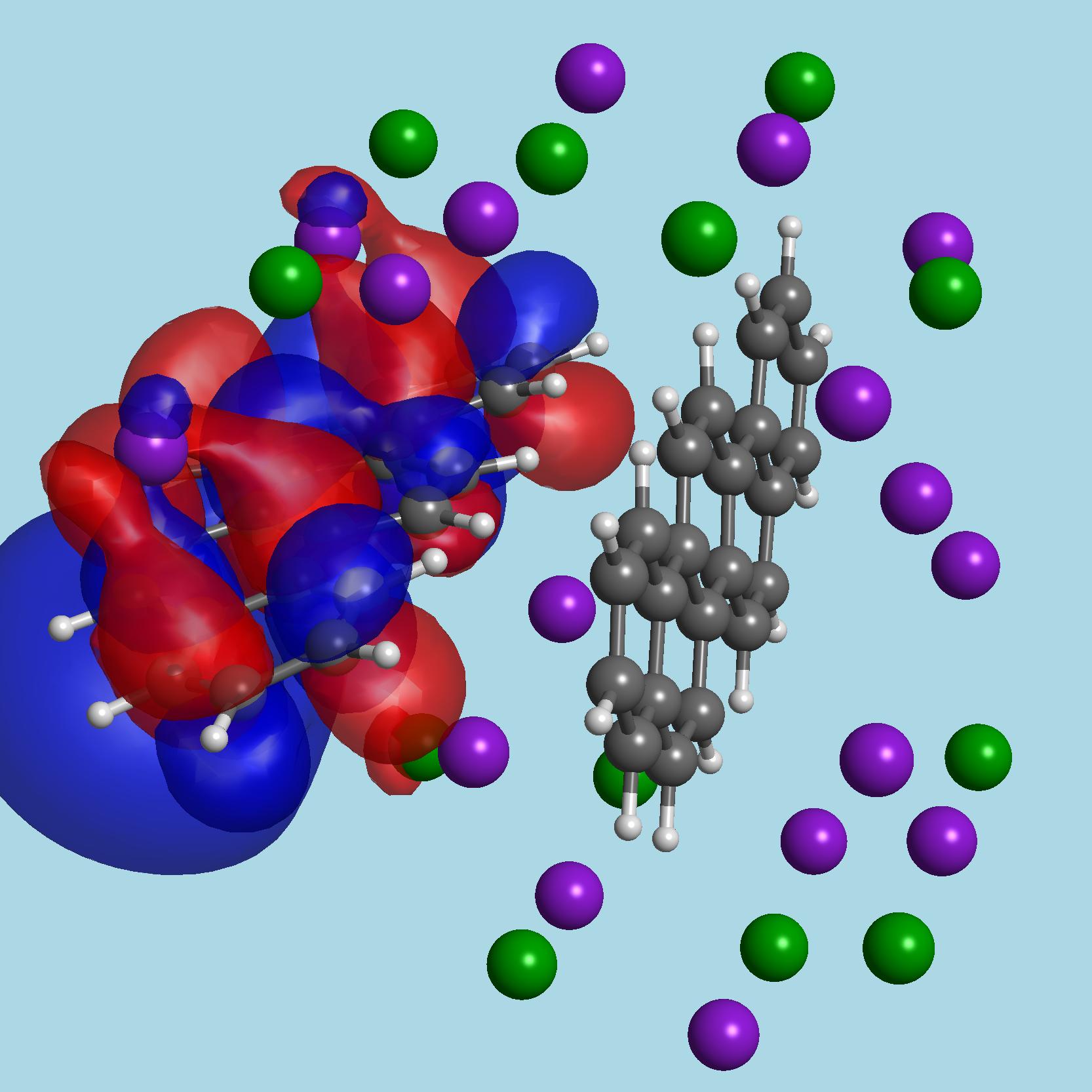}\\
\end{array}$
\caption{(Color online) Higher molecular orbitals (valence electrons) of
a Cl$_{12}$K$_{18}$picene$_2$ neutral cluster designed to
simulate the electronic occupation of crystalline K$_3$picene.
Four electrons occupy MOs based on the LUMOs of isolated picene
molecules while the last two electrons go to a symmetric
combination of potassium $4s$ orbitals (shown in the top
right panel).
}
\label{fig:Cl12K18picene12}
\end{figure}

\subsection{Cl$_{12}$K$_{18}$picene$_2$ neutral cluster}

A graphical representation
of the MOs corresponding to this ternary cluster is shown
in Fig. \ref{fig:Cl12K18picene12}.
The number of valence electrons in the cluster is easily
obtained in this case.
It is simply the number of K atoms minus
the number of chlorine atoms (six for the present composition).
As happened in simpler binary clusters, valence electrons do not flow completely
to unoccupied molecular orbitals of picene molecules. In fact,
four electrons do populate picene LUMOs but the next two go to a
symmetric combination of $4s$ K orbitals. Therefore, the number of
chlorine atoms should be reduced to get the correct K-picene
hybridization. Since this calculation would require a completely new cluster and
the subsequent very tedious geometrical optimization of the structure
and the effects are better described by binary clusters, we have not followed
this line of research any longer.

\begin{figure}
$\begin{array}{cc}
\includegraphics[width=0.49\columnwidth]{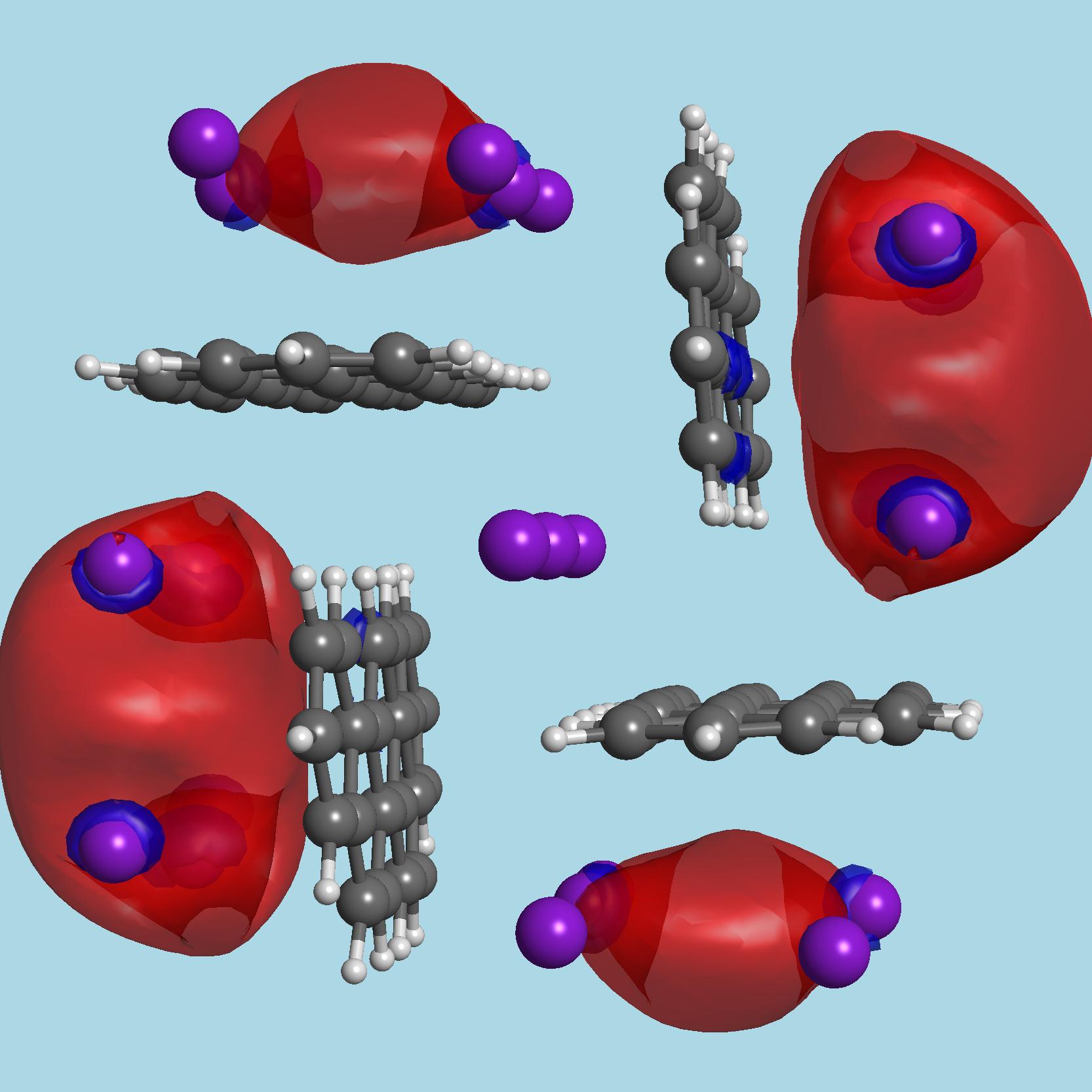}&
\includegraphics[width=0.49\columnwidth]{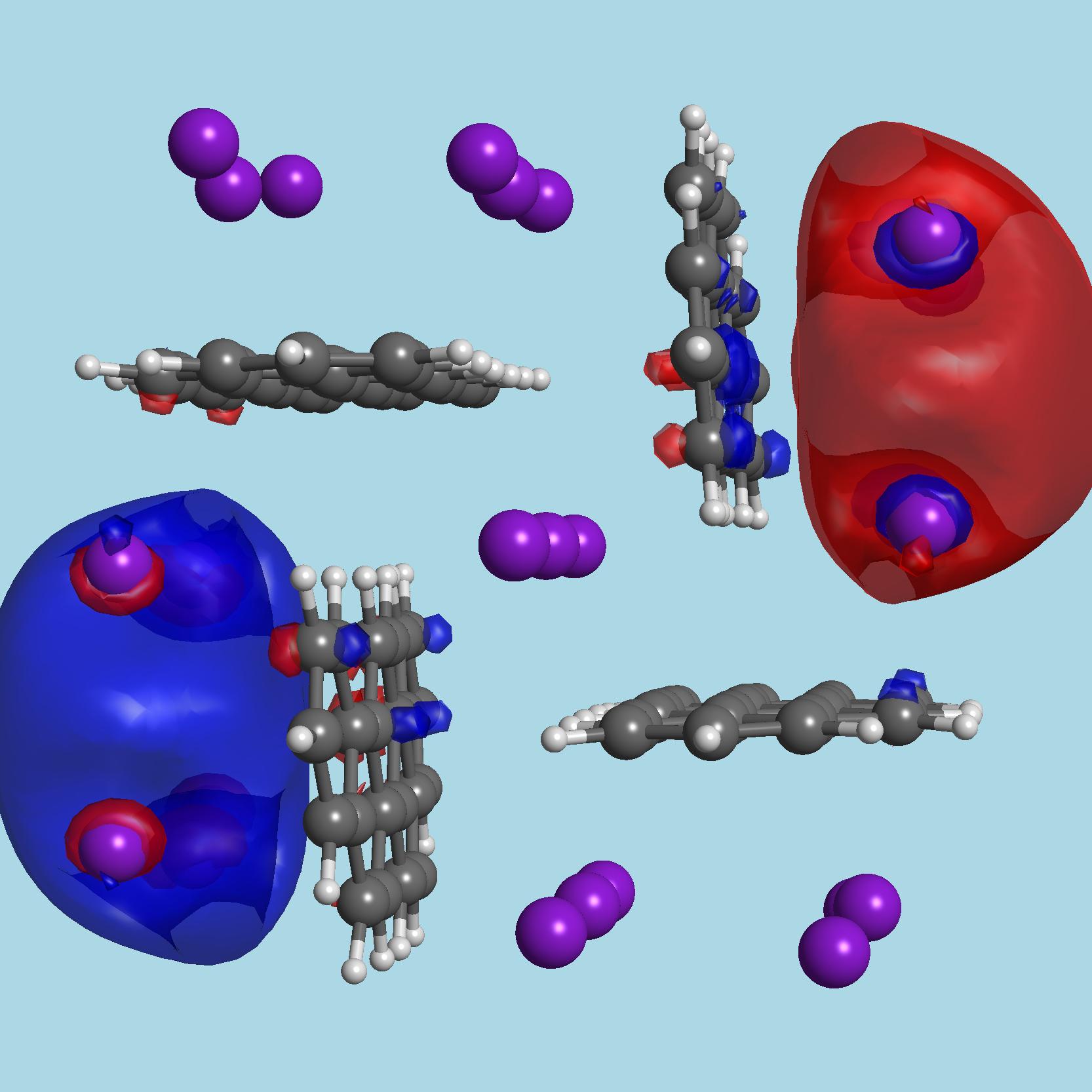}\\
\includegraphics[width=0.49\columnwidth]{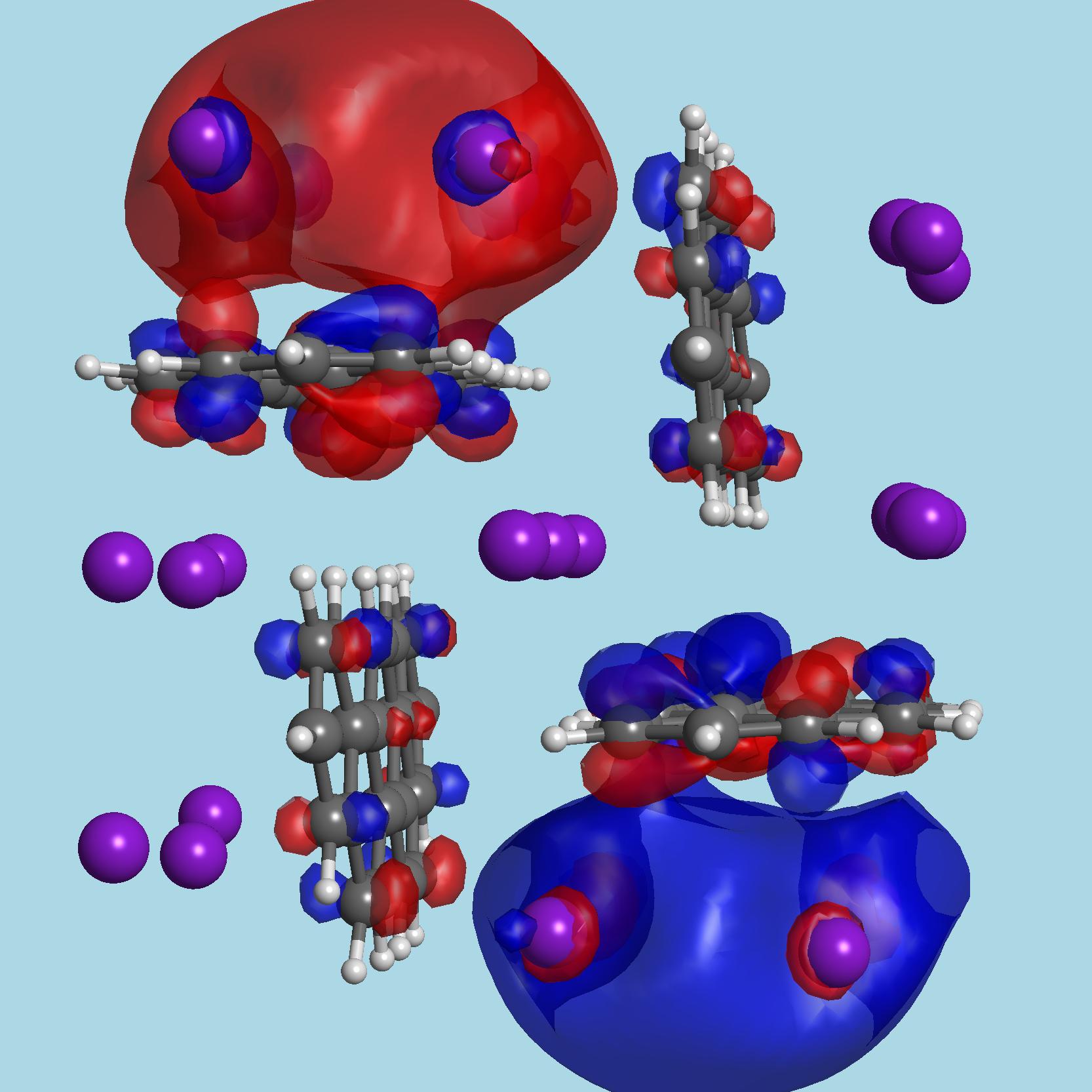}&
\includegraphics[width=0.49\columnwidth]{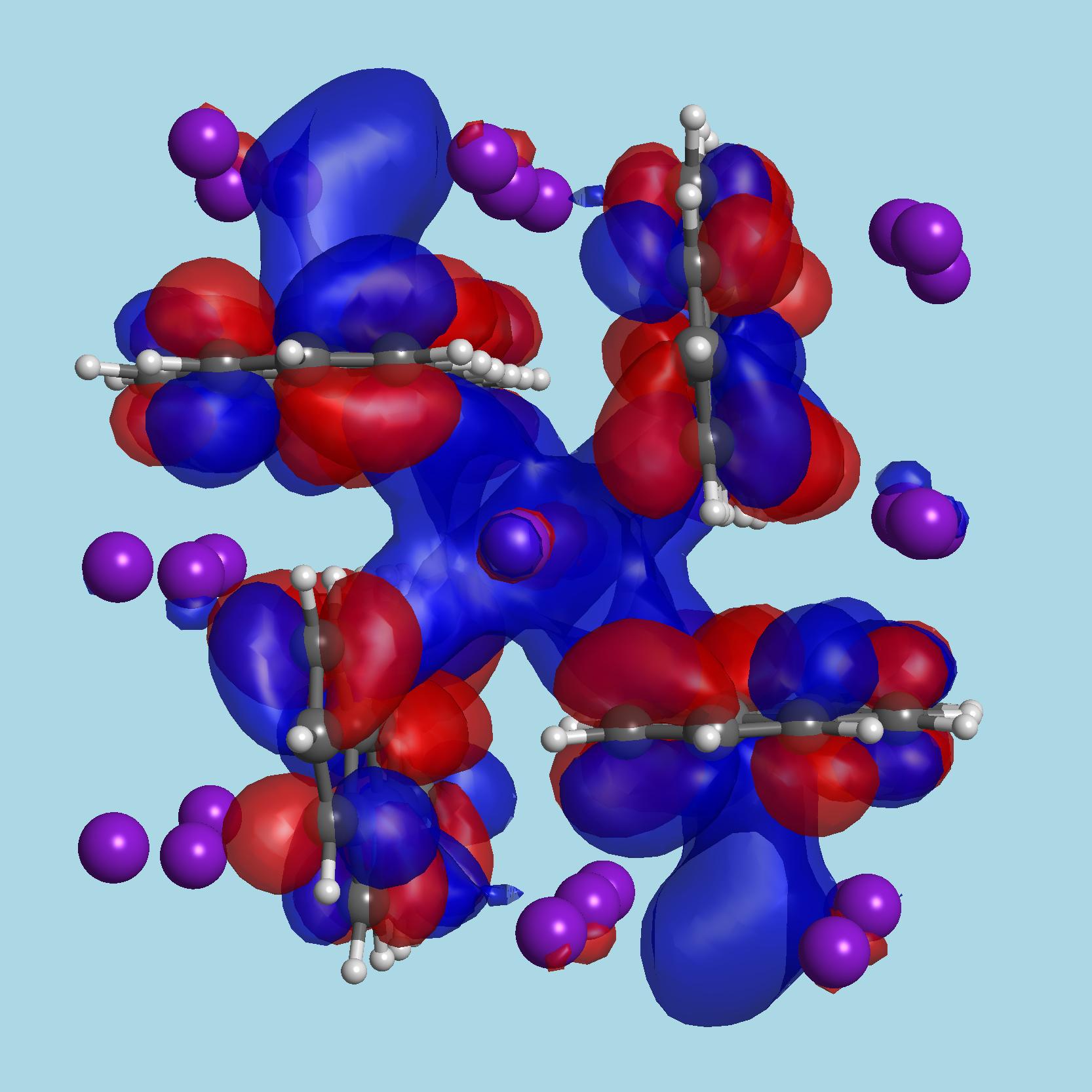}\\
\includegraphics[width=0.49\columnwidth]{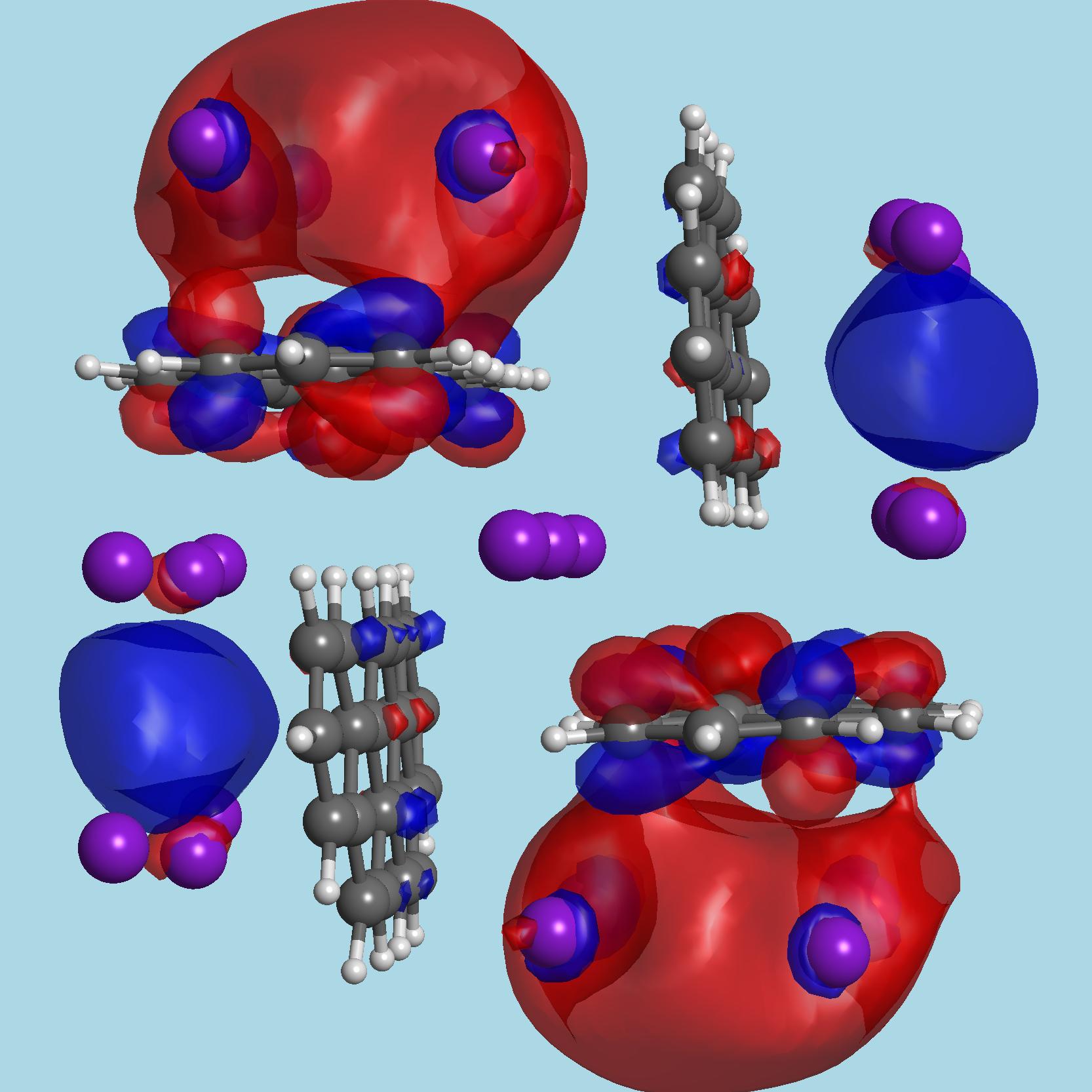}&
\includegraphics[width=0.49\columnwidth]{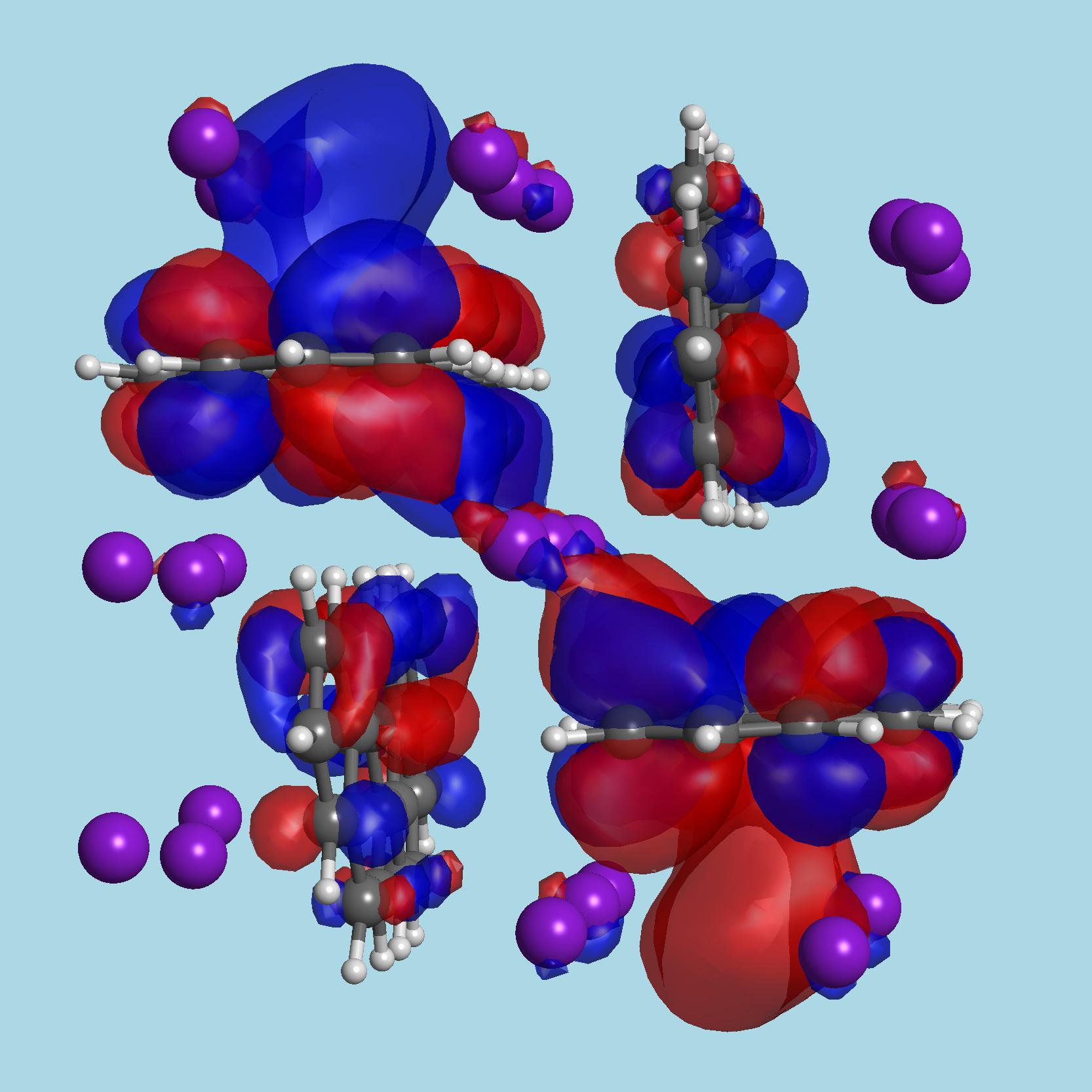}\\
\includegraphics[width=0.49\columnwidth]{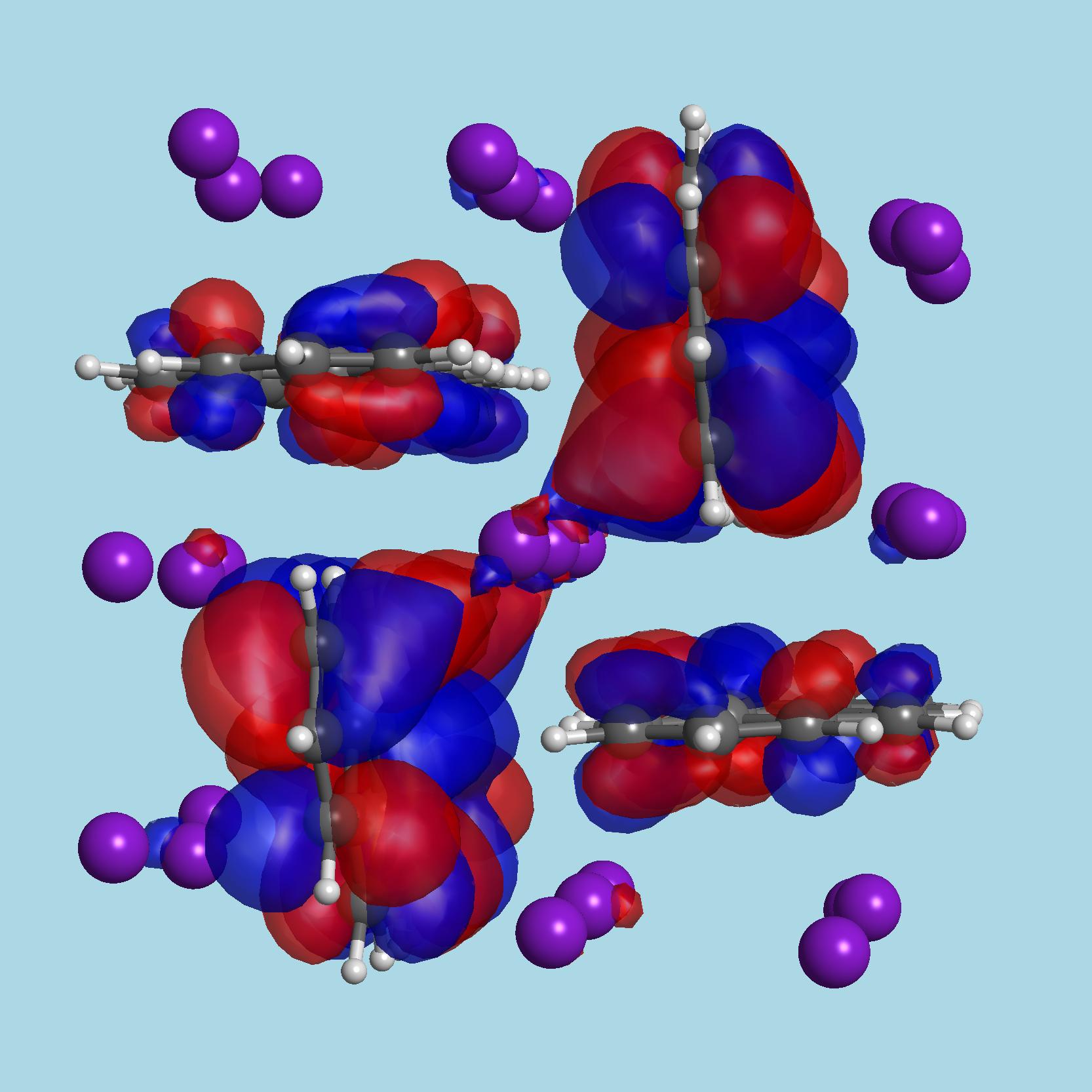}&
\includegraphics[width=0.49\columnwidth]{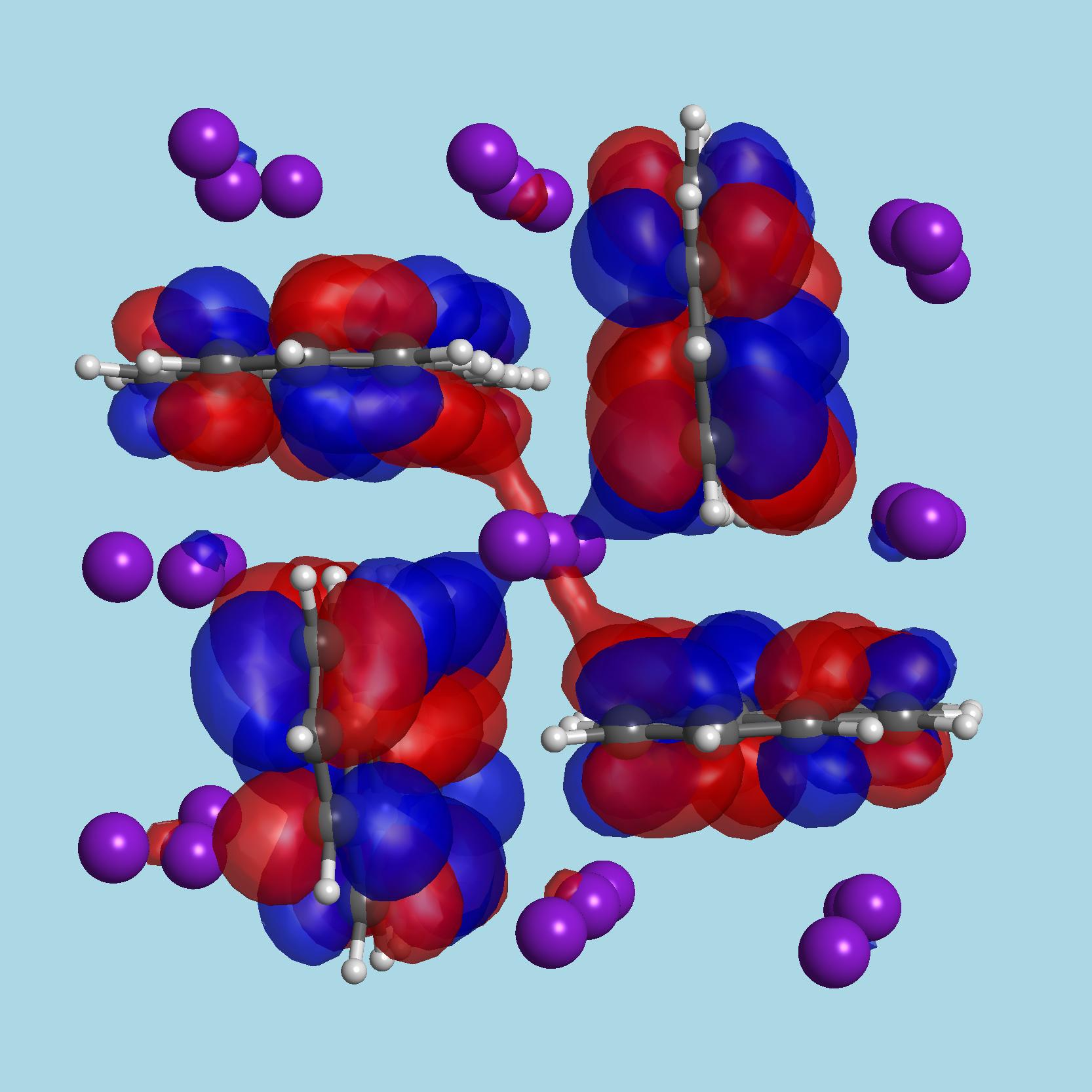}\\
\end{array}$
\caption{(Color online) Most relevant molecular orbitals of
the ground state (a spin heptuplet) obtained for the K$_{27}$picene$_4$$^{13+}$ cluster.
The eight deeper valence electrons going to combinations of
picene LUMOs are not shown. The six remaining valence
electrons singly occupy the first six orbitals shown
in the Figure (following again the numbering
convention stated in the caption of Fig. \ref{fig:clusters}).
The last two orbitals completing the LUMO+1--LUMO+2 shell are empty.
}
\label{fig:K27picene4}
\end{figure}

\subsection{Four picene molecules modeling K$_3$picene}

The DFT optimized geometry obtained for this cluster is shown
in Fig. \ref {fig:clusters} (bottom right panel).
An extra $C_{2h}$ symmetry group was assumed that is
not shared by the crystalline structure. It makes geometry optimization
faster and allows an easier analysis of its electronic structure.
Following previous methodology, fourteen valence electrons have been
considered in this case after geometrical optimization of the cluster
anion (just one extra electron to close shells). Ten of them are
on picene molecules and the rest on potassium atoms. 
Fig. \ref{fig:K27picene4} shows the relevant MOs corresponding to a spin
heptuplet which gives the lowest B3LYP DFT energy (see Table I for
details). There are four electrons occupying symmetric combinations of
$4s$ orbitals on potassium groups and two electrons that are mostly
located on picene described by a particular combination
of picene LUMO+1 and LUMO+2 and some $4s$ K orbitals. The last two
electrons together with
eight electrons on picene LUMOs give a total number of
ten valence electrons on picene molecules.
Additional orbitals that are very close in energy
are also shown in the figure.

\begin{table*}
\caption{Energy of the lower spin excitations of doped and undoped picene clusters. The spin degeneracy
of initial and final states is given in parentheses.}
\begin{ruledtabular}
\begin{tabular}{cccccc}
System & First & Second & Third & Fourth & Fifth \\
\hline
Picene        &    2.23 eV (1$\to$3)   &     4.80 eV (1$\to$5)     & & & \\
Picene$_2$    &    2.21 eV (1$\to$3)   &     4.48 eV (1$\to$5)     & & & \\
Picene$_4$    &    2.15 eV (1$\to$3)   &     4.31 eV (1$\to$5)     & & & \\
\hline
K$_{12}$picene$^{8+}$    &   0.082 eV (3$\to$1)  &  0.179 eV (3$\to$5)  &  3.23 eV (3$\to$7) & & \\
K$_{18}$picene$_2$$^{10+}$ & 0.197 eV (3$\to$5)  &  0.311 eV (3$\to$1)  &  0.726 eV (3$\to$7)  &   0.671 eV (3$\to$9)  &  3.83  eV (3$\to$11) \\
K$_{27}$picene$_4$$^{13+}$ & 0.043 eV (7$\to$5)  &  0.301 eV (7$\to$3)  &  0.358 eV (7$\to$9)  &   0.627 eV (7$\to$11) &  0.685 eV (7$\to$1)  \\
\end{tabular}
\end{ruledtabular}
\end{table*}

When the results for the single-determinantal electronic structure of this
system are analyzed together with excitation energies given in Table I and
the experience gained by the CI analysis of the simplest cluster,
the main message of our work is obtained.
Namely, after doping picene molecules with two electrons
their reduced electronegativity is not enough to gain the extra electron
offered by potassium atoms. This last electron should be described as
belonging simultaneously to PAH molecules {\em and} alkali atoms.
However hopping between LUMO+1 or LUMO+2 and $4s$ K orbitals is very small
mainly due to the rich structure of picene molecular orbitals that
prevents the existence of a naive chemical bond. Therefore, we prefer to say
that the electron can hope between apparently separated K groups just because
the intermediate picene molecule is able to transfer the electron from
one group to the other\cite{IUPAC}.  
It can be said alternatively using a band theory
language that picene LUMO+1 and LUMO+2 become somewhat
hybridized with the K $4s$ state giving rise to a very narrow band
that is partially occupied (one-fourth if two bands are entangled but
one-sixth if three bands are coupled). In fact,
an independent estimation of the hopping energy provides
a value below one tenth of an eV resulting in an estimated bandwidth
of a few tenths of eV. Since it is well known that
only half-filled Hubbard models on bipartite lattices lead to an
unquestionable insulator when correlation effects dominate, our
results strongly suggest that the main reason for the metallic
behavior of K$_3$picene compounds is the small occupation of the
hybridized band. We explore in the next section the behavior of the simplest interaction
model describing this material. Good approximate solutions can be obtained
for larger systems. Results enrich the image obtain for small {\rm all-electron}
systems and give further support for a metallic behavior of
K-intercalated picene even in the presence of very strong electronic correlations.   

Let us finish this subsection commenting Table I. It collects spin excitation energies
for the three K-picene clusters shown in Fig. \ref{fig:clusters} {\em plus} the
clusters produced removing K atoms. We wished to obtain additional electronic
excitations for fixed spin multiplicities but this is a difficult effort in
computational chemistry. Nevertheless, an increased wealth of excitations are
obtained for the interaction model in the next section. The results for undoped cluster
undoubtedly indicate the existence of a gap of about 2.2 eV in the electronic spectrum
of picene molecules. Spin excitations are produced promoting one paired
core electron to an unoccupied level with a large energy cost.
Flipping an additional electron to produce a spin quintuplet implies two electrons in
excited states and more than twice the gap energy.
On the other hand, spin excitation
energies are much smaller for the doped clusters pointing therefore to a dense
spectrum of many-body states. Also notice that excitation energy becomes above 3 eV
when an electron on a picene LUMO needs to be promoted to a higher energy level
to form the spin multiplet. In short, we attribute the existence of
small energy electronic excitations
of doped clusters to the richness of LUMO+1--LUMO+2--$4s$(K) configurations
participating in accurate many-body wavefunctions.

\subsection{K charge in a large K$_3$picene cluster}

\begin{figure}
$\begin{array}{c}
\includegraphics[width=0.98\columnwidth]{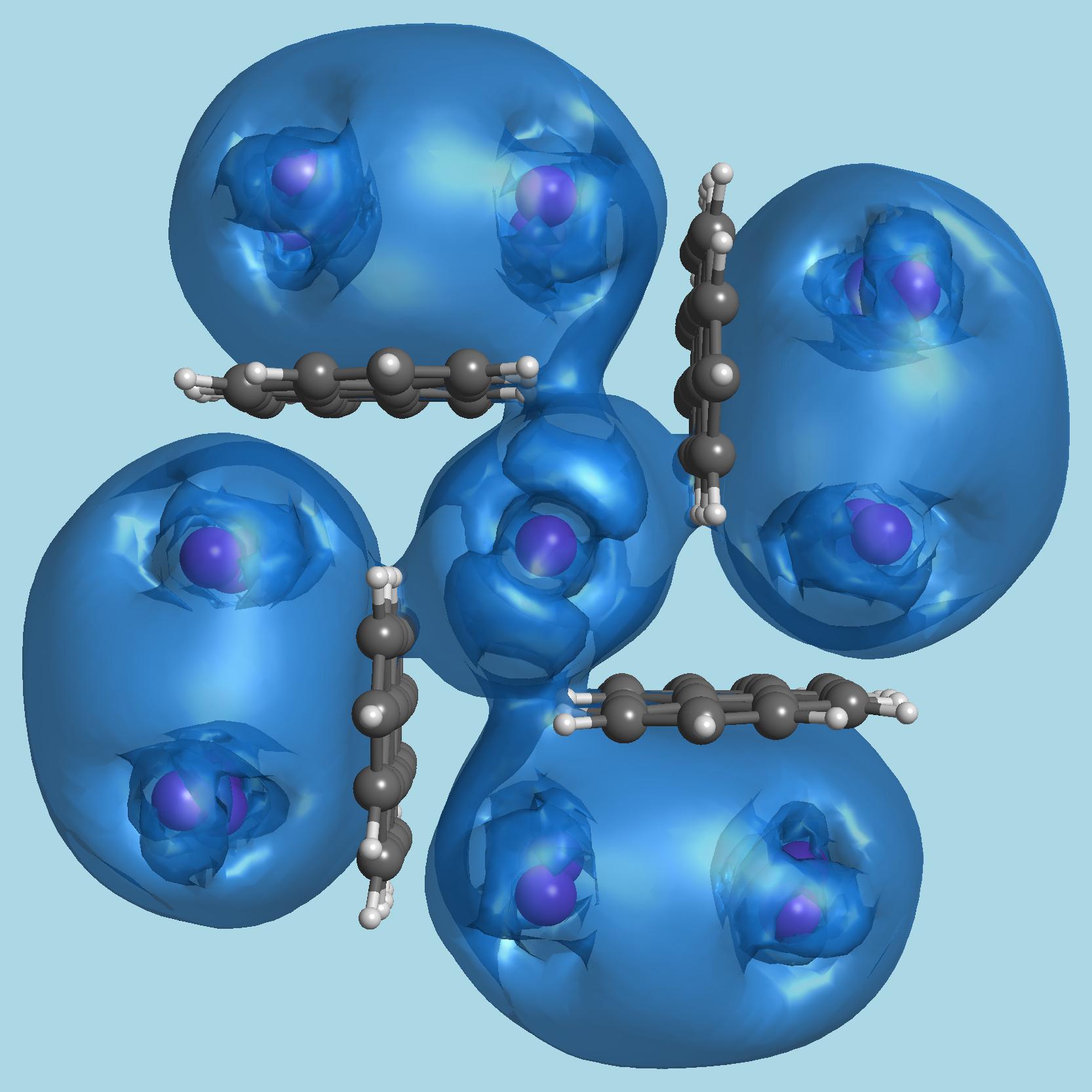} \\
\includegraphics[width=0.98\columnwidth]{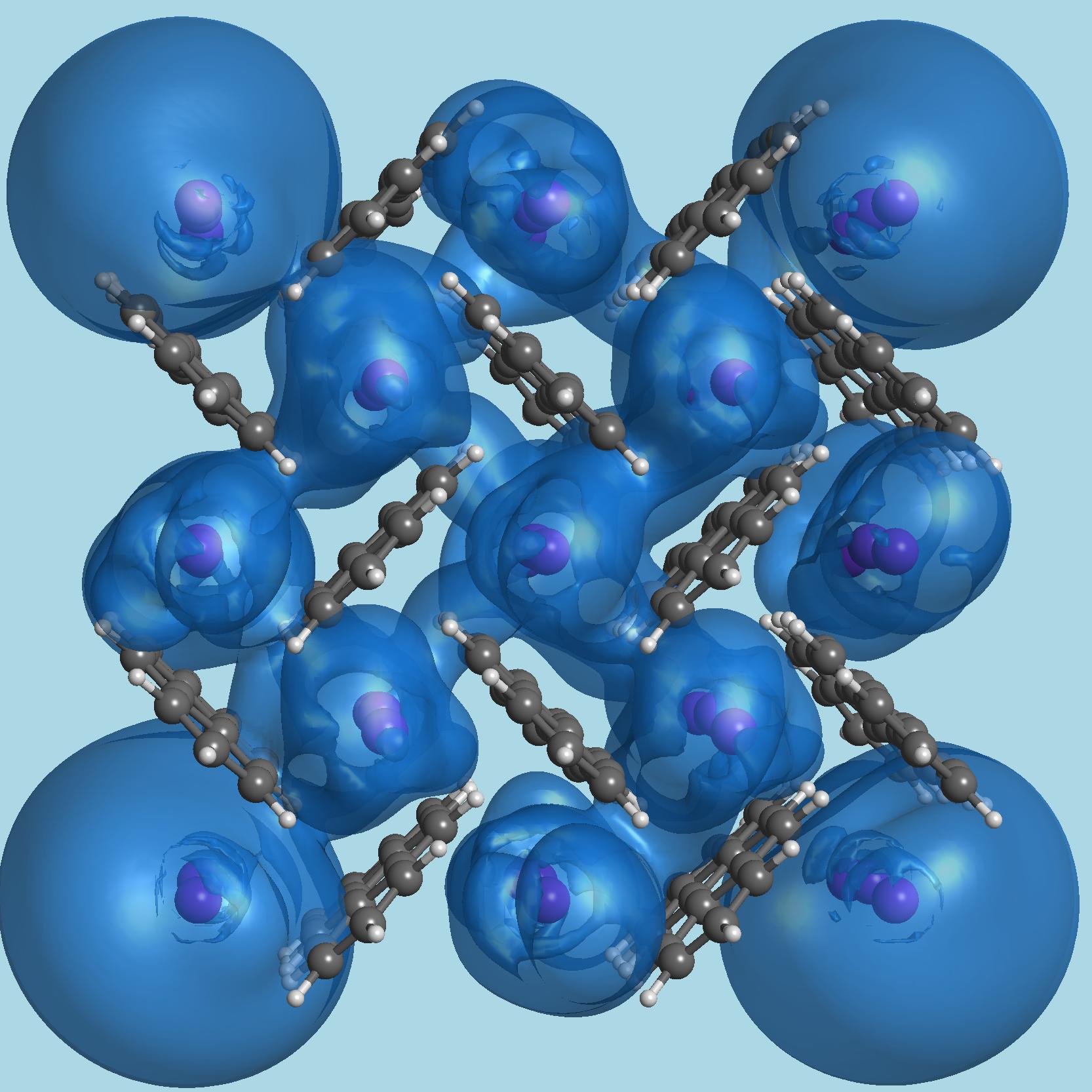} \\
\end{array}$
\caption{(Color online) Charge density corresponding to valence electrons projected onto K orbitals.
The distribution in the cluster studied in subsection IIID (upper panel) is compared with
the distribution of a much larger cluster (lower panel).}
\label{fig:SUPER}
\end{figure}

A large K$_3^{3+}$(K$_3$picene)$_{12}$ cluster taken from our crystalline structure
(Ref. \onlinecite{pedro2011}) has been solved
using B3LYP functional and a PC1 gaussian basis\cite{jensen}.
Cluster and crystal are isoelectronic in this case.
The total number of valence electrons is 36. These valence electrons populate picene molecular orbitals
that are unoccupied in the pristine material and K orbitals.
Fig. \ref{fig:SUPER} shows the charge distribution associated to K atoms. At the corners it is similar to the one found on
smaller clusters for external K$_6$ groups while follows the one found for the confined K$_3$ group of
{K$_{27}$picene$_4$}$^{13+}$ cation for all potassium groups surrounded by four picene molecules.
Charge density isosurface has been plotted at a small value of $3 \times {10}^{-5}$ e \AA$^{-3}$ to show
that overlap between K and picene orbitals should exist although its value will be presumably tiny.

\section{Many-body model Hamiltonian}

\subsection{Charge gap}

A simple way to infer values for the energies due to interaction among electrons
in a cluster is
changing the total number of electrons in it. Table II gives the values obtained
for our clusters not only when K is present but also when potassium is completely
removed from the system. The importance of size effects can be correctly estimated
by the comparison of the values obtained in doped systems with the ones
got for pure systems.
Actually, we calculate the charge gap $E_{N+1}+E_{N-1}-2 E_N$ for a cluster
with $N$ electrons in the reference electronic state. It measures the insulating
character of the system. It is obviously always finite for clusters but can
go to zero in infinite systems. Values for undoped clusters (fourth column)
are much larger that for K-doped systems (second column) and both show a clear
diminution with size. We know that values for the last column should tend to
the semiconducting gap of picene (about 2.2 eV) but it is impossible to say what
will happen with the charge gap for crystalline K$_3$picene although an
eventual size scaling towards zero could occur.
Here it is a good place to remember some works emphasizing
that quarter-filled bands can give metallic conduction in spite
of showing a finite charge gap\cite{model1,model2,model3}. 
Moreover, small energy spin excitations as those included in Table I for
doped clusters point to metallic behavior not only because they could approach zero for larger systems
but also because they imply charge displacement\cite{sutileza}.

\begin{table}
\caption{Charge gap calculated for both doped and undoped picene clusters using expression
$E_{N+1}+E_{N-1}-2 E_N$ where $E_N$ is the total DFT energy of the reference system 
($N$ electrons).}
\begin{ruledtabular}
\begin{tabular}{cc|cc}
Doped cluster & Gap energy & Undoped cluster & Gap energy \\
\hline
K$_{12}$picene$^{8+}$      &  2.72 eV & Picene     &  6.10 eV \\
K$_{18}$picene$_2$$^{10+}$ &  2.37 eV & Picene$_2$ &  5.34 eV \\
K$_{27}$picene$_4$$^{13+}$ &  1.83 eV & Picene$_4$ &  4.86 eV \\
\end{tabular}
\end{ruledtabular}
\end{table}

\subsection{Two band model}

Here we propose a minimalist interaction model Hamiltonian for K$_3$picene
crystal. It is based in the main result of our quantum chemistry calculations
that show an important hybridization between picene and potassium orbitals
in spite of the very small direct coupling. A straightforward calculation of
the Hubbard U (intrasite Coulomb repulsion) gives between three and four eV
both on picene and on a group of three aligned K atoms. The precise value
depends on details as occupation, geometry and others.
Coulomb repulsion for separate electrons
can be estimated from the Coulomb law. The last ingredient of the model
is hopping from picene LUMO+1 or LUMO+2 to bonding $4s$ combinations. It
is necessarily small since unoccupied $\pi$ states of picene show alternating
positive and negative lobes while $4s$ combinations show a very mild spatial
dependence. In order to get a concrete value for the hopping we have used
a simple multiconfigurational calculation based on the two orbitals that
show the main coupling in the simplest system
(the K$_{12}$picene$^{9+}$ cluster presented in Subsection IIIA).


Parameters of the interacting model are obtained from a series of simple CI calculations
of the K$_{12}$picene cluster shown at the upper left corner of Fig. \ref{fig:clusters}.
Many-body states are calculated for
electronic populations that go from two to six valence electrons. Two electrons occupy
an A$_2$ molecular orbital (very similar to the first one of Fig. \ref{fig:K12picene})
while the rest populate two B$_2$ molecular orbitals, that is,
orbitals showing the symmetry of the second
natural orbital of Fig. \ref{fig:K12picene} but localized on picene or K atoms at the mean-field level.
The energy of the state with two electrons on the slightly modified picene LUMO (A$_2$ molecular
orbital) is considered as the reference.  
Both CI energies and wavefunctions are taken into account to fit the parameters
of an interacting model now including only two states. One corresponds to picene and the second
to K. We start fitting the case of one extra electron which is described by the
following Hamiltonian matrix: 

\begin{displaymath}
{\mathbf H}_{1e} =
\left( \begin{array}{cc}
\epsilon_1  &  t  \\
t  &  \epsilon_2  \\
\end{array} \right) \; ,
\end{displaymath}

\noindent
where $\epsilon_i$ are the state levels and $t$ the hopping between them.
The following values are obtained:
$ \epsilon_1 =$ -24.00 eV, $ \epsilon_2 =$ -23.95 eV and $t =$ -0.048 eV.
Eigenvector amplitudes are $(0.846252, 0.532783)$ and $(-0.532783, 0.846252)$
for both {\em ab initio} and model calculations.

The Hamiltonian matrix describing two extra valence electrons is:

\begin{displaymath}
{\mathbf H}_{2e} =
\left( \begin{array}{cccc}
2 \epsilon_1 + U_1  &  0  &  t  &  t \\
0  &  2 \epsilon_2 + U_2  &  t  &  t \\
t  &  t  &  \epsilon_1 + \epsilon_2 + V  &  0 \\
t  &  t  &  0  & \epsilon_1 + \epsilon_2 + V  \\
\end{array} \right) \; ,
\end{displaymath}

\noindent
where $U_i$ are on-site Coulomb repulsion energies while $V$ gives the
Coulomb repulsion energy between picene centered and K centered molecular orbitals. The triplet
state (0,0,1,-1) allows the determination of $V$. We get $V =$ 2.99 eV.
The on-site Coulomb parameters are obtained diagonalizing the Hamiltonian matrix describing
three extra valence electrons:

\begin{displaymath}
{\mathbf H}_{3e} =
\left( \begin{array}{cc}
2 \epsilon_1 + \epsilon_2 + U_1 + 2 V  &  t  \\
t  &  \epsilon_1 + 2 \epsilon_2 + U_2 + 2 V  \\
\end{array} \right) \; .
\end{displaymath}

\noindent
Values of $U_1 =$ 3.31 eV and $U_2 =$ 4.81 eV are obtained.
Eigenvector amplitudes are $(0.994478, -0.104948)$ and $(0.104948, 0.994478)$
for both {\em ab initio} and model calculations. Finally, the case of four
extra electrons filling completely both B$_2$ molecular orbitals is described by:

\begin{displaymath}
{\mathbf H}_{4e} =
\left( \begin{array}{c}
2 \epsilon_1 + 2 \epsilon_2 + U_1 + U_2 + 4 V  \\
\end{array} \right) \; .
\end{displaymath}

\noindent
Singlet states for two electrons are obtained by diagonalization of ${\mathbf H}_{2e}$.
They are given in the Table III compiling all {\em ab initio} and model energies.
Agreement is good for the three singlets and also for the completely filled system.

\begin{table*}
\caption{Energies of a two-state interacting model as a function of electron
population. Energies of the two last columns are referred to the empty state.}
\begin{ruledtabular}
\begin{tabular}{cccc}
Number of extra electrons & CI eigenenergy (Hartree)  & {\em Ab initio} energy (eV) & Model energy (eV) \\
\hline
0     & -7988.5174 &     0        &     0      \\
1     & -7989.3958 &   -24.03     &   -24.03   \\
1     & -7989.3919 &   -23.92     &   -23.92   \\
2     & -7990.1656 &   -44.98     &   -44.98   \\
2 (triplet)    & -7990.1650 &   -44.96     &   -44.96   \\
2     & -7990.1535 &   -44.65     &   -44.67   \\
2     & -7990.0973 &   -43.12     &   -43.10   \\
3     & -7990.8159 &   -62.67     &   -62.67   \\
3     & -7990.7581 &   -61.10     &   -61.10   \\
4     & -7991.2991 &   -75.82     &   -75.82   \\
\end{tabular}
\end{ruledtabular}
\end{table*}

\begin{figure}
\includegraphics[width=0.98\columnwidth]{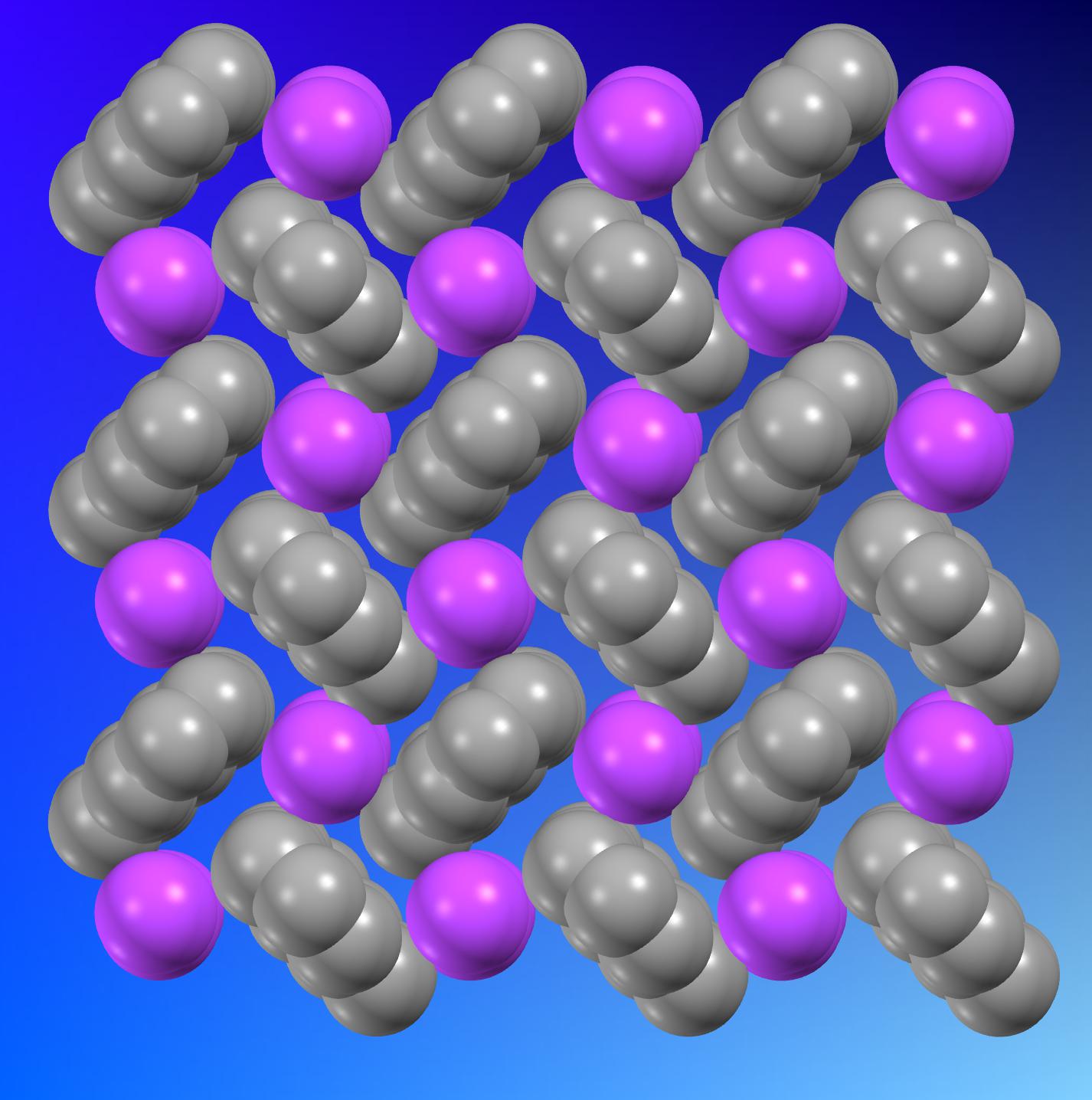}
\caption{(Color online) $6 \times 6$ cluster illustrating our interaction
model study. The large size of atoms (80 \% of the Van der Waals radii) tries 
to simulate spatial extension of involved orbitals: LUMO+1 or LUMO+2 on
picene molecules and bonding $4s$ combinations on the K rows. We ignore direct
hopping between picene orbitals owing to their symmetry. Consequently, hopping
is only considered between LUMO+1s (or LUMO+2s)
and their neighboring four rows of K atoms.
In the same way, hopping of K $4s$ combinations is limited
to their four neighboring picene orbitals.
Although all hopping energies are taken equal for simplicity, signs take into
account the $\pi$-character of picene orbitals.
}
\label{fig:cluster6x6}
\end{figure}

Finally, a hopping of -0.05 eV correctly fits {\em ab initio} results.
On the other hand, alternative values for the Coulomb integrals are obtained
as a byproduct. Notice
that $U/t \sim 60-80$ which implies a strongly correlated electronic system
that would lead to an insulating behavior in any bipartite lattice.
As a further simplifying detail,
we will study an isolated layer of K$_3$picene ignoring possible
three-dimensional effects. Fig. \ref{fig:cluster6x6} shows a $6 \times 6$ cluster
of the model in which both the perspective and the large size of atoms helps
to imagine the relevant orbitals.
The bidimensional array is formed by alternating
picene and K sites in an underlying square lattice.
Nevertheless, the model differs from a simple interaction model on the
square lattice because
the sign of the hopping energy should be carefully
chosen to take into account both the $\pi$-character of PAH molecular orbitals
and the particular staggering of picene molecules
in the herringbone structure.

The model Hamiltonian contains
a non-interacting part $\hat H_{0}$ and a term that incorporates the
electron-electron interactions $\hat H_{I}$:

\begin{equation}
{\hat H}  = {\hat H_{0}} + {\hat H_{I}}\;.
\label{eq:H}
\end{equation}

\noindent
The non-interacting term is written as:

\begin{equation}
{\hat H_{0}}  =  \sum_{i\sigma} \epsilon_{i}
{\hat c}^{\dagger}_{i\sigma} {\hat c}_{i\sigma} +
\sum_{ij\sigma} t_{ij} {\hat c}^{\dagger}_{i\sigma} {\hat c}_{j\sigma}\;,
\label{eq:H_{0}}
\end{equation}

\noindent
where the operator ${\hat c}^{\dagger}_{i\sigma}$ creates an electron
at site $i$ with spin $\sigma$, $\epsilon_{i}$
is the energy of the orbital at site $i$,
and $t_{ij}$ is the hopping between orbitals at sites $i$ and $j$, respectively.

The interacting part is given by:

\begin{equation}
{\hat H_{I}}  =
U\sum_{i}{\hat n}_{i\uparrow}{\hat n}_{i\downarrow}+{\frac{1}{2}}
\sum_{ij; i \neq j} V_{ij} ({\hat n}_{i}-1/2) ({\hat n}_{j}-1/2)\;,
\label{eq:H_{I}}
\end{equation}

\noindent
where $U$ is the on-site Coulomb repulsion and $V_{ij}$ is
the inter-site Coulomb repulsion given by:

\begin{equation}
V_{ij}=U \left[1+\left(\frac{U}{e^2/d_{ij}}\right)^2\right]^{-1/2}\;,
\end{equation}

\noindent
being $d_{ij}$ the distance between $i$ and $j$ sites.

\noindent
The electronic density operator for spin $\sigma$ at site $i$ is:

\begin{equation}
{\hat n}_{i\sigma}= {\hat c}^{\dagger}_{i\sigma} {\hat c}_{i\sigma}\;,
\label{eq:density}
\end{equation}

\noindent
whereas the operator giving the total electron density at
the same site is:

\begin{equation}
{\hat n}_{i}= {\hat n}_{i\uparrow} + {\hat n}_{i\downarrow}\;.
\end{equation}

\noindent
Let us remind that we use
$\epsilon_{i} = 0$ eV, $t_{ij} = \pm 0.05$ eV for nearest neighbors
and $U = 4.0$ eV in an underlying square lattice of lattice constant
equal to $3.5$ \AA ~as a minimum model describing
resonance between picene and K valence electrons.
Sometimes we have used an ionic variation of this model ($\epsilon_{i}=0$
on picene sites but $\epsilon_{i}=0.4$ eV on K sites) to check that
qualitative results do not depend on the oversimplification of the
proposed model.

The model Hamiltonian is solved using a development of Lanczos method that
follows Quantum Chemistry standards. A mean-field approach is used to
obtain uncorrelated HF molecular orbitals. Afterward, the interaction model
is rewritten in this new basis and CI employed. The crucial point is that
selection of configurations is not based on the occupation of the
initial wavefunction but generated from it by successive applications
of the many-body Hamiltonian (keep in mind that the use of the Hamiltonian operator
to increase the Hilbert space dimension constitutes the essence of
the Lanczos method). 
New configurations are kept only when their participation
in the ground state is relevant, that is, when its amplitude is above
some carefully chosen threshold. If the threshold is too small, the number
of configurations exceeds computational capabilities. On the other hand, if
the threshold is too large only a few configurations are generated and
correlation is poorly included. Typically, no more than several hundreds of
thousands of configurations are used to describe the correlated system.
However, the way they have been selected produces a final result showing very
good quality. Although our method is applied to a model Hamiltonian and a
purely numerical algorithm is used for the selection of configurations,
we believe that it shares some similarities with a well documented set of
wise criteria of CI selection used for {\em ab initio} calculations\cite{CI}.
Further details on the whole Lanczos procedure can be found in recent
papers\cite{verges2012,SpinStates}. 

\begin{table}
\caption{Exact results for the interaction model on a small $4 \times 4$
cluster. Second column corresponds to equal levels on all sites while
the third column gives the values for the more general ionic version.
A large number of digits is necessary to show the very small
energy differences between the lowest many-body states.}
\begin{ruledtabular}
\begin{tabular}{ccc}
 State & E$_{\rm covalent}$ (eV) & E$_{\rm ionic}$ (eV) \\ 
\hline
 Ground state            & -7.568838 & -7.534038 \\
 First excited singlet   & -7.568685 & -7.533984 \\
 Second excited singlet  & -7.568658 & -7.533972 \\
 First triplet           & -7.568823 & -7.534021 \\
 First quintuplet        & -7.568812 & -7.534003 \\
 Anion GS                & -6.275474 & -5.617117 \\
 Cation GS               & -6.258459 & -5.930108 \\
\end{tabular}
\end{ruledtabular}
\end{table}

\subsection{
Some exact results for a small $4 \times 4$ cluster
}

\begin{figure}
\includegraphics[width=0.98\columnwidth]{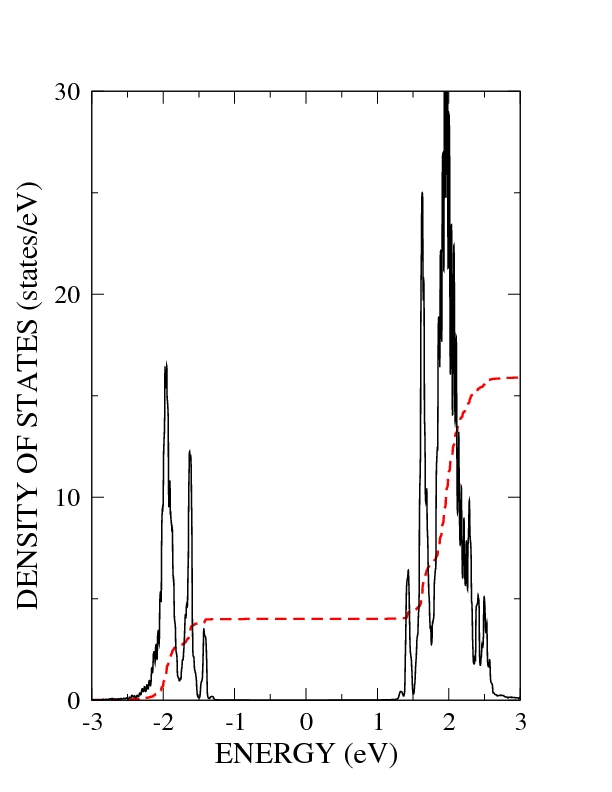}
\caption{(Color online) The density of states (DOS) of a $4 \times 4$ cluster with eight
electrons described by the model Hamiltonian is shown in black.
The lower band is fully
occupied (four spin up electrons and four spin down electrons) while
the upper one is empty. Integrated occupation per spin is given by the
dashed red line (the numerical scale is valid for the integrated DOS).}
\label{fig:DOS0}
\end{figure}

The density of states of a $4 \times 4$ cluster has been calculated within
Lanczos formalism for the minimalist model Hamiltonian described in the precedent subsection.
The simplest version with $\epsilon_i=0$ for all sites $i$ was chosen.
The result is given in Fig. \ref{fig:DOS0}. It shows two Hubbard bands centered on $\pm \frac{1}{2} U$
and separated by an energy equal to the charge gap obtained from the data in Table IV.
The lower band is completely filled with eight electrons (four of each spin) while
the upper one is empty.

Table IV collects some exact results for the same $4 \times 4$ cluster using
in this case both covalent and ionic flavors of the lattice model.
Anion and cation energies together
with the ground state energy allow obtaining the charge gap.
Values of 2.60 and 3.52 eV result for covalent and ionic models,
respectively. They overestimate values in Table II pointing to an excessively
large $U$ value. On the other hand, excitation energies are much smaller than
those given in Table I for {\rm all-electron} clusters. This fact points again
to an unrealistic dominance of interaction over hopping.
Therefore, it should be inferred that the many-body physics contained in the
quantum-chemistry results is not completely reproduced by the simplest model
that needs at least an {\em ad hoc} modulation of the $U/t$ ratio.
In any case, diminishing $U/t$ is always possible and will lead to a
reinforcement of the signals pointing to a metallic behavior of the system.

\subsection{
HOMO-HOMO correlation function.
}

Lanczos methods not only allow the calculation of the correlated many-body
wavefunction but also obtaining dynamical properties like response functions
or correlation functions. After the selection of some convenient operator $\hat O$,
it is applied to the ground state at $t$=0 and propagated up to time $t$ using the
exact many-body Hamiltonian. After Fourier transforming the time variable,
correlation is given by:

\begin{equation}
C(\omega)=-\frac{1}{\pi} Im<GS|\hat O^{\dagger} 
          (\omega+E_{\rm GS}+i\epsilon-\hat H)^{-1} \hat O|GS>\; ,
\label{eq:correlation}
\end{equation}


\noindent
where $|GS>$ is the many-body ground state and $E_{\rm GS}$ its energy.
The propagator
can be written in tridiagonal form using a Lanczos representation
of Hilbert space (see section IIA2 of Dagotto's review\cite{dagotto}).
Here, we will follow the time response of the system
to the creation of an electron-hole
pair to reveal low energy excitations that does not change the total number of particles
in the system. Usually, Lanczos formalism is used to get one-electron
Green functions in which
case one electron (or hole) is created at $t$=0 and its time evolution followed
before destruction at time $t$. 
In these cases, time propagation occurs via many-body states containing one
more (or one less) particle and the correlated density of states is obtained.
Since we know that a large charge gap will separate empty from occupied states
due to size effects, we prefer to infer information from two-particle
electronic excitations using correlation functions.
In fact, the time evolution of a pair excitation obtained by annihilation of
one electron in an antibonding combination of HOMOs and creation in the bonding
combination has been calculated, i.e.,

\begin{equation}
\hat O = {\hat c}^{\dagger}_{\rm homo1} {\hat c}_{\rm homo2}\; ,
\label{eq:correlation2}
\end{equation}

\noindent
where homo1 stands for the bonding linear combination of the two HOMOs that
produces the preliminary mean-field calculation whereas homo2 stands
for the antibonding linear combination of the same MOs.

\begin{figure}
\includegraphics[width=1.2\columnwidth]{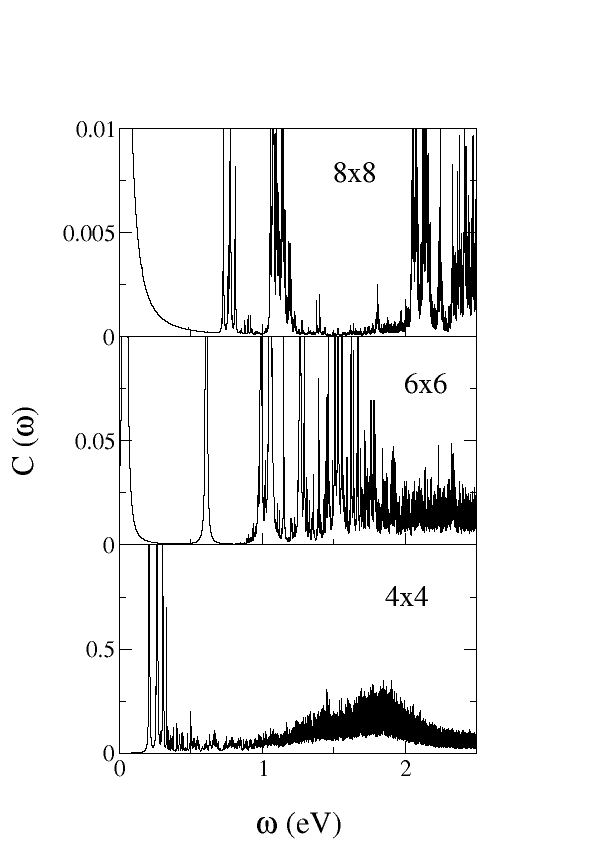}
\caption{
HOMO-HOMO correlation function defined by
Eqs.(\ref{eq:correlation}-\ref{eq:correlation2}) as a function
of the cluster size. The time evolution of an HOMO-HOMO electron-hole
pair is followed being the pair formed by bonding and antibonding linear
combinations of the two HOMOs obtained in the initial mean-field
approximation of the correlated state.
}
\label{fig:DOS}
\end{figure}

Results corresponding to $4 \times 4$, $6 \times 6$ and, $8 \times8$
clusters are shown in Fig. \ref{fig:DOS}.
We have checked that mean-field HOMOs are strongly localized on distant lattice
sites. Therefore, $C(\omega)$ can be seen as the time response to a large dipole
perturbation. It starts displaying
several peaks at small energies (some tenths of eV) that are
followed by an incoherent part at larger energies (one or two eV).
In this way, the existence of small energy electronic excitations involving
charge displacement is proved for relatively large clusters.
A more detailed exposition of Lanczos methods and results is postponed for
a future publication.

\section{Concluding Remarks}

We want to finish the presentation of our basically computational work with
a few remarks based on the more robust results.
i) Doping of picene with potassium completely populates the LUMO of picene producing
a band that is inert from the point of view of electronic transport.
ii) The third electron of the K$_3$ group is not transferred to picene LUMO+1 or
LUMO+2 but shared among PAH molecule and K atoms in spite of the very small
hopping between the corresponding orbitals.
iii) Since hopping is small but interaction is large, the correct description of the
electronic structure requires a many-body formulation.
iv) Both quantum-chemistry results for small {\rm all-electron} clusters and results
obtained from a minimalist interaction model coincide in getting small energy
electronic excitations that imply charge displacement.  
v) These results open the way to explain metallic behavior in spite of the strong
interaction shown by electrons at the Fermi level.

Comparison of our results with previous theoretical contributions allows to find
some similarities
with Kim {\em et al.} works that assign metallic behavior to complex Fermi surfaces formed
by several bands\cite{kim2011,kim2013} but also important disagreement
with conclusions derived by
other authors based on the large gap shown by the spectral function of
the strongly correlated electronic system (see Fig. 3 of Ref. \onlinecite{ruff2013}).
Notice that our interaction model is similar to theirs
(although we are coupling K and picene
bands and our $U$ value is sensibly larger), and therefore,
our quarter-filled two-band model produces a spectral
function that resembles the one shown at the top of the mentioned figure
(Fig. \ref{fig:DOS0} for a $4 \times 4$ cluster).
Nevertheless, we argue that low energy two-particle excitations are responsible
for transport instead of more common one-particle excitations. As already
mentioned above, a detailed discussion of the interaction model will be presented
elsewhere.

Once a metallic phase is made plausible for K$_3$picene compound,
the possibility of superconductivity cannot be excluded
from a purely theoretical point of view. In this context both many-body
effects and electron-phonon interactions would be
important\cite{giovannetti2011,verges2012,zhong2013,ruff2013,kim2013,SC1,SC2,SC3}.
Finally, let us remind that after the seminal work by
Mitsuhashi {\em et al.}\cite{mitsu2010} showing the existence of
superconducting phases for potassium-intercalated picene, additional
experimental evidence points to a metallic behavior of K$_3$picene.
See, for example, the observation of
orbital hybridization between picene (C $2p$) and potassium (K $4s$) near the Fermi energy
obtained by Yamane and Kosugi\cite{yamane2012} using soft x--ray spectroscopies
or the zero resistivity measurement by Teranishi {\em et al.}\cite{ZeroResistivity}.
Moreover, metallic states in K-intercalated picene films on
graphite have been observed very recently\cite{okazaki2013}.


\begin{acknowledgments}
Financial support by the Spanish MICINN
(MAT2011-26534, CTQ2007-65218, CSD2007-6, FIS2012-33521,
FIS2012-35880 and CTQ2011-24165)
and the Universidad de Alicante is gratefully acknowledged.
We also acknowledge support from the DGUI
of the Comunidad de Madrid under the R\&D Program of activities
MODELICO-CM/S2009ESP-1691.
\end{acknowledgments}


\begin{thebibliography}{99}
\bibitem{core}
Let us count electrons starting from neutral picene molecules and
completely ionized potassium atoms (K$^+$). From here on, we will refer to
the electrons in this initial electronic configuration
as core electrons because their wavefunctions are very close
to the ones corresponding to isolated molecules or ions. The
electrons that are relevant for the transport properties of the compound
will be called valence electrons (exactly three per picene molecule
assuming a stoichiometric K$_3$picene compound).
\bibitem{sutileza}
It could be argued that a half-filled Hubbard model for large values of $U$ also shows
small energy spin excitations. In this case small energy excitations and insulating
behavior appear simultaneously. Nevertheless, the situation seems to be completely different
at smaller fillings. In these cases spin excitations are coupled to charge movement and
metallic conduction is possible. Later we will use the proposed model to show some
illuminating examples. Coming back to the binary compound here studied, it is clear
that small energy spin excitations are associated to charge transfer from K to picene orbitals
making a small energy charge transport quite plausible.
\end{thebibliography}
\end{document}